\documentclass[12pt]{article}
\pdfoutput=1	

\usepackage{geometry}
\usepackage{physics}
\usepackage{graphicx}				 	
\usepackage{amssymb,bbm}
\usepackage{amsmath}
\usepackage{jheppub}
\usepackage{tikz}
\usepackage{xcolor}
\usepackage[makeroom]{cancel}

\newcommand{\nc}{\newcommand}
\nc{\rnc}{\renewcommand }

\rnc{\a}{\alpha}
\rnc{\b}{\beta}
\nc{\g}{\gamma}
\rnc{\d}{\delta}
\nc{\e}{\epsilon}
\nc{\z}{\zeta}
\nc{\f}{\phi}
\nc{\m}{\mu}
\nc{\n}{\nu}
\rnc{\r}{\rho}
\rnc{\k}{\kappa}
\rnc{\l}{\lambda}
\nc{\s}{\sigma}
\rnc{\t}{\tau}
\nc{\w}{\omega}
\nc{\vp}{\varphi}

\nc{\A}{\Alpha}
\nc{\B}{\Beta}
\nc{\G}{\Gamma}
\nc{\D}{\Delta}
\rnc{\S}{\Sigma}
\rnc{\P}{\Pi}

\nc{\bD}{\mathbb{D}}
\nc{\bQ}{\mathbb{Q}}
\nc{\bX}{\mathbb{X}}
\nc{\bY}{\mathbb{Y}}
\nc{\bS}{\mathbb{S}}
\nc{\bR}{\mathbb{R}}
\nc{\bC}{\mathbb{C}}
\nc{\bW}{\mathbb{W}}
\nc{\bF}{\mathbb{F}}
\nc{\mc}[1]{\mathcal{#1}}
\nc{\mcs}{\mathbf{S}}

\nc{\mm}{\mathbf{m}}
\nc{\nn}{\mathbf{n}}
\nc{\pp}{\mathbf{p}}
\nc{\na}{\nabla}
\nc{\pa}{\partial}

\nc{\MM}{\mathbf{M}}
\nc{\NN}{\mathbf{N}}
\nc{\PP}{\mathbf{P}}

\nc{\beq}{\begin{equation}}
\nc{\eeq}{\end{equation}}
\nc{\bea}{\begin{align}}
\nc{\eea}{\end{align}}

\nc{\rcom}{\bf\large \color{red}}
\nc{\bcom}{\bf\large \color{blue}}

\title{3D Dualities and Supersymmetry Enhancement from Domain Walls}
\author[1]{Martin Ro\v cek,}
\author[1]{Konstantinos Roumpedakis,}
\author[2]{Sahand Seifnashri.}
\affiliation[1]{C. N. Yang Institute for Theoretical Physics\\
Stony Brook University\\
Stony Brook, NY 11794-3840, USA.}
\affiliation[2]{Simons Center for Geometry and Physics\\
Stony Brook University\\
Stony Brook, NY 11794-3636, USA. \\[5mm]}

\abstract{
We test recently proposed IR dualities and supersymmetry enhancement by studying the supersymmetry on domain walls. In the $SU(3)$ Wess-Zumino model studied in \cite{Gaiotto:2018yjh,Benini:2018bhk}, we show that domain walls exhibit supersymmetry enhancement. This model was conjectured to be dual to an $\mc N=2$ abelian gauge theory. We show that domain walls on the gauge theory side are consistent with the proposed duality, as they are described by the same effective theory on the wall. In \cite{Fazzi:2018rkr}, a third model was conjectured to be dual to the same IR theory. We study the phases and domain walls of this model and we show that they also agree. We then consider the analogous $SU(5)$ Wess-Zumino model, and study its mass deformations and phases. We argue that even though one might expect supersymmetry enhancement in this model as well, the analysis of its domain walls shows that there is none. Finally, we study the $\mc{N}=2$ model in \cite{Gang:2018huc} which was conjectured to have $\mc{N}=4$ supersymmetry in the IR. In this case we don't see the supersymmetry enhancement on the domain wall; however, we argue that half-BPS domain walls of the $\mc{N}=2$ algebra are quarter-BPS of the $\mc{N}=4$ algebra. This is then in agreement with the conjectured enhancement, even though it does not show that it takes place.}

\emailAdd{~\\~~martin.rocek@stonybrook.edu~\\~ 
konstantinos.roumpedakis@stonybrook.edu~\\~
sahand.seifnashri@stonybrook.edu}

\preprint{YITP-SB-19-8}
\begin{document}
\maketitle
\section{Introduction}

Three dimensional quantum field theories (QFT) exhibit a variety of infrared (IR) phases with interesting features. Although it is hard to determine the phases of a general QFT, supersymmetry is often a useful tool for extracting such information. This work focuses on two interesting aspects of IR behavior in three dimensional supersymmetric theories: IR dualities and supersymmetry enhancement. 

Three dimensional dualities have a long history with the well-known examples of particle-vortex \cite{Peskin:1977kp,Dasgupta:1981zz}, vector-vector  \cite{Deser:1982vy, Karlhede:1986qf, Karlhede:1986qd, Banerjee:2000hs}, scalar-vector \cite{Polyakov:1988md} and non-supersymmetric level-rank \cite{Naculich:1990pa,Mlawer:1990uv,Naculich:2007nc} dualities.
In the last decade, following the original papers \cite{Seiberg:1994pq,Intriligator:1995au} on IR duality in four dimensions, a plethora of new dualities in three dimensions have been proposed. Ranging from various non-supersymmetric examples \cite{Aharony:2013dha,Aharony:2013kma,Seiberg:2016gmd,Hsin:2016blu,Aharony:2016jvv,Komargodski:2017keh,Gomis:2017ixy,Choi:2018tuh} to supersymmetric cases \cite{Aharony:1997gp,Bashmakov:2018wts,Gaiotto:2018yjh,Benini:2018bhk,Armoni:2017jkl,Armoni:2018ahv, Choi:2018ohn,Bashmakov:2018ghn, Benini:2018umh}, the idea of duality has shed a lot of light on the infrared structure of three-dimensional QFTs. 
 
For applications to real world physics, supersymmetry is usually regarded as a ultraviolet (UV) symmetry which is broken in the IR. However, some modern research \cite{Lee:2006if,Yu:2010zv,Ponte:2012ru,Grover:2013rc} in condensed matter physics has exploited the prospect of supersymmetry as an emergent symmetry of physical systems at low energies. Along these lines, another possibility, which is more tractable to analyze, is the case of supersymmetry enhancement \cite{Ritz:2004mp,Bashkirov:2010kz,Maruyoshi:2016tqk,Maruyoshi:2016aim,Agarwal:2016pjo,Agarwal:2017roi,Gang:2018huc,Giacomelli:2018ziv,Evtikhiev:2017heo}.

The purpose of this paper is to test various recently proposed IR dualities and the idea of supersymmetry enhancement by analyzing the effective theories on the world-volume of domain walls.  Domain walls in supersymmetric gauge theories were first studied in \cite{Dvali:1996xe,Kovner:1997ca,Chibisov:1997rc,Acharya:2001dz,Eto:2018nbt}, and reviewed in \cite{Eto:2006pg, Shifman:2007ce, Shifman:2009zz}.

For testing IR dualities the main idea we use is to deform both sides of the proposed duality until multiple degenerate vacua exist, and then consider domain walls interpolating between such vacua. These classical objects carry charges that extend the super-Poincar\'e algebra and therefore are stable configurations. The goal then is to show that on the two-dimensional world-volume of the wall, the effective theory for localized zero modes is the same on both sides of the duality. 

For exploring supersymmetry enhancement we analyze the amount of supersymmetry on the domain-wall theory. Classical solitonic solutions of the equation of motion, such as domain walls in supersymetric theories, often have the profound feature of being annihilated by a subset of the supercharges \cite{Witten:1978mh}. In three dimensions, typically, they are half-BPS objects and the two dimensional world-volume theory inherits half of the supersymmetry of the parent bulk theory. However, by explicitly studying the localized zero-modes on the wall, in some cases one can prove that the two-dimensional theory has actually more that half of the bulk supersymmetry, suggesting that the parent three-dimensional theory has enhanced supersymmetry in the IR.

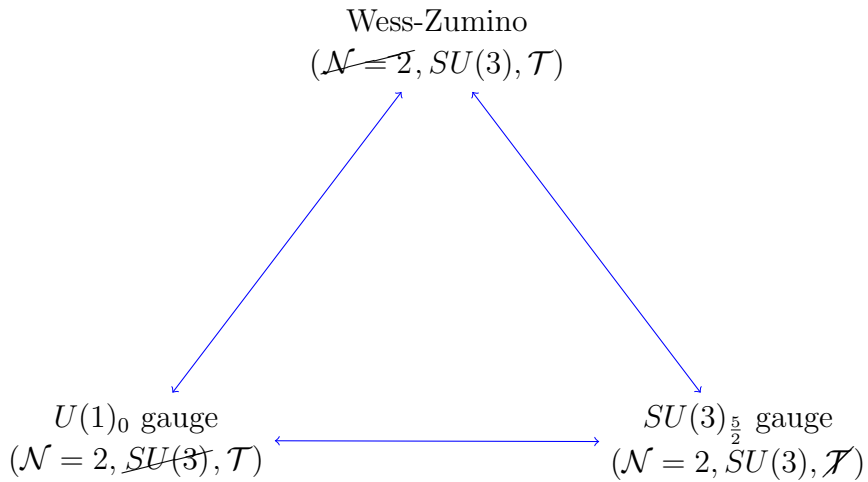
\begin{figure}[ht]
\centering
\begin{tikzpicture}[scale=2] 
    \draw
    (0,0) node(1)[align=center,above] {Wess-Zumino \\ $(\cancel{\mc{N}=2}, SU(3), \mc{T})$}
   (-2,-2) node(2)[align=center,below] {$U(1)_0$ gauge \\ $(\mc{N}=2, \cancel{SU(3)}, \mc{T})$}
   (2,-2) node(3)[align=center,below] {$SU(3)_\frac{5}{2}$ gauge  \\ $(\mc{N}=2, SU(3), \cancel{\mc{T}})$}
 -- cycle;
 \draw[<->,blue] (1) -- (2);
 \draw[<->,blue] (2) -- (3);
 \draw[<->,blue] (1) -- (3);
    \end{tikzpicture}
    \caption{The models depicted in this figure are dual to each other. For each model we indicate which symmetries are manifest in the UV action and which are not (crossed out).}
    \label{fig:1}
\end{figure}

One of the main results of this work is the test of dualities between three different theories proposed in \cite{Gaiotto:2018yjh,Benini:2018bhk,Fazzi:2018rkr}. The three models participating in the duality web are depicted in figure \ref{fig:1}.

\noindent None of the UV theories manifestly exhibits all of the three global symmetries that the IR theory is believed to have, namely $\mc{N}=2$ supersymmetry, $SU(3)$ flavor symmetry, and time reversal invariance. We consider domain walls in all  three theories and we show that in each case, the effective theory on domain walls is a chiral $(2,0)$ theory with target space ${\bR}\times S^1$, while for anti-domain walls the theory is a chiral $(0,2)$ with the same target space.

We then consider a Wess-Zumino model with a cubic superpotential where the fields transform in the adjoint representation of an $SU(5)$ flavor symmetry. As in the case of $SU(3)$ model we break the flavor symmetry by giving masses to the scalars. The moduli spaces of vacua for these deformed theories are always complex Grassmannians and therefore K{\" a}hler manifolds. This might lead one to believe that there is supersymmetry enhancement for these deformed theories and for the IR fixed point of the undeformed theory, with the extra supercharges coming from the complex structure of the moduli space. However, further  analysis of domain walls shows that actually supersymmetry is not enhanced: For generic linear deformations we have degenerate vacua with domain walls that interpolate between them; investigating the effective theory on the walls, we show that there exist domain-wall solutions with a single bosonic zero mode and therefore, only $(1,0)$ supersymmetry. This rules out the possibility of enhancement.\footnote{One might imagine that domain walls in the dual theory could preserve only $\frac14$ of the supersymmetry; however, we argue the multiplet structure of these theories prevents that scenario.}

The third main result of this paper is a consistency check in the presence of domain walls for the recently conjectured supersymmetry enhancement to $\mc{N}=4$ for an $\mc{N}=2$ abelian gauge theory with Chern-Simons level at $k=-3/2$ with one chiral multiplet of charge one. We add an FI term which leads to two degenerate vacua and we consider domain walls. We argue that the solutions of the classical BPS equations have only one modulus which not only seems to contradict the enhancement but also the bulk $\mc{N}=2$ supersymmetry. We resolve the contradiction in two steps. First, we study the infinity coupling limit of the BPS equation and we show that in this regime on one side of the domain wall we get an $U(1)_{-1}$ gauge theory. Although the vacuum of this theory is trivial, in the presence of a domain-wall boundary it gives rise to a chiral zero mode as in the Hall-effect \cite{Tong:2016kpv}. Furthermore, we address the apparent contradiction with the $\mc{N}=4$ enhancement. We argue that the FI deformation of the $\mc{N}=2$ action corresponds to a new kind of mass deformation that was pointed out in \cite{Cordova:2016xhm}. These deformations change the supersymmetric algebra by a term involving the $\mc{R}$-symmetry current. We then show that domain walls in the $\mc{N}=4$ deformed theory are actually quarter-BPS and therefore, they lead to two unbroken supercharges in agreement with the enhancement.

To find these results, we develop various tools: we write down the defining equations for the 
$\mc{N}=1$ $\mcs$-multiplet in three dimension, generalizing the $\mc{N}=2$ results of \cite{Dumitrescu:2011iu,Komargodski:2010rb}, and we derive explicit expressions for the Wess-Zumino model as well as abelian gauge theories. We also derive the explicit expression for the 
$\mcs$-multiplet of an arbitrary $\mc{N}=2$ non-abelian gauge theory with a superpotential in three dimension. We calculate the brane-charges from the $\mcs$-multiplet and obtain the tension of the domain walls that these theories support. We use Morse theory to analyze the solutions of the BPS equations that we find in various models. Use of Morse theory in the study of BPS domain walls can also be found in \cite{Bachas:2000dx,Gauntlett:2000ib, Eto:2005wf}. In the pure Wess-Zumino model, the superpotential defines a Morse function and we can use the results of Morse theory directly to analyze the BPS equations; in other cases, we construct the Morse function by hand.

The outline of the paper is as follows. In section \ref{sec:2}, we review the $\mc{N}=2$ $\mcs$-multiplet of \cite{Dumitrescu:2011iu} and give explicit expressions for abelian and non-abelian gauge theories. In section \ref{sec:3}, we present the defining equations for the  $\mc{N}=1$ $\mcs$-multiplet and give explicit expressions in several examples. In section \ref{sec:DM}, we discuss the kinematics of the BPS-saturated domain walls in $\mc{N}=1$ theories from the supersymmetry algebra. In Section \ref{sec:5}, we present a new check for the duality of the three theories in figure \ref{fig:1} by studying the domain walls in these models. In section \ref{sec:6} we study the $\mc{N}=1$ $SU(5)$ WZ model and discuss its possible supersymmetry enhancement in the infrared. In section \ref{sec:2 to 4}, we study the $\mc{N}=2$ to $\mc{N}=4$ supersymmetry enhancement of the model presented in \cite{Gang:2018huc} by considering domain walls. In appendix \ref{sec:conventions}, we summarize our conventions about $\mc{N}=1$ and $\mc{N}=2$ supersymmetry and gauge-covaraint superderivatives. In appendix \ref{sec:sugra}, we review $\mc{N}=1$ supergravity and in appendix \ref{sec:Universal mass} we discuss an example of a relevant deformation in the stress-tensor multiplet of 
$\mc{N}=4$ SCFTs. 
\section{\texorpdfstring{$\mcs$}{S}-multiplet for \texorpdfstring{$\mc{N}=2$}{N=2} theories \label{sec:2}} 
\subsection{Defining equations} \label{sec:def for N=2}
In supersymmetric theories, operators organize themselves into multiplets of the Super-Poincar\'e algebra. Generally, representations containing conserved currents belong to multiplets obeying some shortening conditions. In superconformal theories, short supermultiplets saturate the unitarity bound \cite{Mack:1975je}, and they have been fully classified \cite{Dolan:2002zh,Cordova:2016emh}. When one deforms a superconformal theory to break conformal symmetry, the short superconformal multiplets combine with other multiples to make a short supermultiplet of the super-Poincar\'e algebra, and the shortening conditions are modified. A universal multiplet that exists in any local supersymmetric theory is the stress-energy multiplet. Two well-known examples of stress-energy multiplets are the Ferrara-Zumino multiplet \cite{Ferrara:1974pz} and the $\mc{R}$-multiplet \cite{Gates:1981yc,Dienes:2009td,Kuzenko:2009ym} in four dimensional $\mc{N}=1$ theories. These two multiplets turn out to be special cases of a more general stress-energy multiplet, referred to as the $\mcs$-multiplet in \cite{Komargodski:2010rb,Kuzenko:2010am, Dumitrescu:2011iu} and elsewhere.

Apart from the energy momentum tensor, the $\mcs$-multiplet also contains the supercurrent as well as any $\mc R$-symmetry currents (which however need not be conserved). In addition, it may also contain additional conserved currents that modify the super-Poincar\'e algebra. These terms can be either central extensions, or brane currents that are may be present in domain-wall backgrounds. The purpose of this section is to review the $\mcs$-multiplet for three-dimensional 
$\mc{N}=2$ theories, and the possible extensions of the $\mc{N}=2$ superalgebra. We summarize our conventions in appendix \ref{sec:conventions}, where we also review several useful identities.

In $\mc{N}=2$ theories, the defining set of equations for the $\mcs$-multiplet \cite{Dumitrescu:2011iu}
are\footnote{We have set the constant $C=\bar{{\bD}}^\a{\bX}_\a=-\bar{{\bD}}^\a{\bY}_\a$
in eq.$\!$ (4.1) of  \cite{Dumitrescu:2011iu} equal to zero. It gives rise to space-filling charges 
which are not of interest for the discussion of domain walls.} 
\bea
& \bar{{\bD}}^\b \mcs_{\a\b}={\bX}_\a+ {\bY}_\a~, \nonumber \\
& \bar{{\bD}}_\a {\bX}_\b=0~, ~\quad {\bD}^\a {\bX}_\a=- \bar{{\bD}}^\a \bar{{\bX}}_\a~, \nonumber \\
&{\bD}_{( \a} {\bY}_{\b)}=0, \quad \bar{{\bD}}^\a {\bY}_\a=0~; \label{N=2 S}
\end{align}
here $\mcs_{\a\b}$ is a real vector superfield, whereas $\bX_\a,\bY_\a$ are complex spinor superfields.

The conserved supercurrent and the energy-momentum tensor are the lowest components of the superfields
\beq
S_{\b\g}{}^\a=i \bD^\a \mcs_{\b\g}-i\d^\a_{(\b} \bar{\bY}_{\g)}~,
\eeq
and
\beq
T_{\a\b,\g\d} =-\frac{1}{8}\left( [\bD_{(\a},\bar{\bD}_{\b)}] \mcs_{\g\d} +[\bD_{(\g},\bar{\bD}_{\d)}] \mcs_{\a\b}\right)-\frac{1}{16}\e_{\a(\g}\e_{\d)\b} \left([\bD_{\e},\bar{\bD}_{\z}] \mcs^{\e\z} + 2\bD^\z \bX_\z\right)   ~,
\eeq
respectively.
As was shown in \cite{Dumitrescu:2011iu}, apart from these two conserved quantities and possibly an 
$\mc R$-symmetry current, the stress-energy multiplet contains three additional tensors given by 
\bea
F_{\a\b,\g\d} & =\frac{i}{16}\left[\e_{\a\g}(\bD_{(\b} \bX_{\d)}+\bar\bD_{(\b} \bar\bX_{\d)})+
 \e_{\b\d}(\bD_{(\a} \bX_{\g)}+\bar\bD_{(\a} \bar\bX_{\g)}) \right]+\left[\a\leftrightarrow\b\right]~,\label{F current}\\
H_{\a\b} & = -\frac{1}{16} \left( \bD_{(\a} \bX_{\b)}-\bar{\bD}_{(\a} \bar{\bX}_{\b)} \right)~,\label{H current} \\
\bar{Y}_{\a\b} & =\frac{i}{2} \bD_{(\a} \bar{\bY}_{\b)}~.
\end{align}
The duals of these tensors are conserved currents.
In vector notation (see Appendix \ref{indices}), the supercurrent and the energy-momentum tensor can be identified as the lowest component of the superfields
\bea
S^\a_{\m}&= i{\bD}^\a \mcs_{\m} - \frac{i}{2} \g_\m^{\a\g} \bar{{\bY}}_{\g}~,\nonumber \\
T_{\m\n}&=-\frac{1}{16} \g_{(\m}^{\a\b} [\bD_{\a},\bar{\bD}_{\b}]\mcs_{\n)} +\frac{1}{8} \eta_{\m\n} \g_\r^{\g\d} {\bD}_\g \bar{{\bD}}_\d\mcs^\r+\frac{1}{8} \eta_{\mu\nu}{\bD}^\a {\bX}_\a~.
\end{align}
while the additional tensors are
\bea
F^{\n \r} & = \frac{i}{32} \e^{\n\r\m} \g_\m^{\a\b} \left( {\bD}_\a {\bX}_\b + \bar{{\bD}}_\a \bar{{\bX}}_\b \right)~, \\
H_\m & = -\frac{1}{32} \g_\m^{\a\b} \left( {\bD}_\a {\bX}_\b - \bar{{\bD}}_\a \bar{{\bX}}_\b \right)~, \\
\bar{Y}_\r & =\frac{i}{4} \g_\r^{\a\b} {\bD}_\a \bar{ {\bY}}_\b~.
\end{align}
Using these expressions one can determine the algebra of two supercharges 
\beq
{\bQ}_\a=\int d^2 x  \;  S_{\a}\!^0{\Big|}~.
\eeq 
The anticommutator of the supercharge with the supercurrent is
\bea
\{ \bar{\bQ}_\a, S_{\m\b}{\big|}\}  = i\bar\bD_\a S_{\m\b}{\Big|}~=~&  \g^\n_{\a\b} \left( 2 T_{\m\n}+ 2 \e_{\m\n\r}  H^\r+i \pa_\n \mcs_\m -i \eta_{\m\n} \pa^\r \mcs_\r\right)\!{\Bigg|}\\
 & +i \e_{\a\b} \e_{\m\n\r} \left( 2 F^{\n\r} + \pa^\n\mcs^\r \right){\Bigg|}~, \\
\{ \bQ_\a, S_{\m\b}|\}  =   i\bD_\a S_{\m\b}{\Big|}~=~& i \e_{\m\n\r} \g^\n_{\a\b}  \bar{Y}^\r{\Big|}~.
\end{align}
The terms proportional to $\mcs_\m$ give rise to Schwinger terms. The other extra currents on the left-hand side extend the super-Poincar\'e algebra to
\bea
\{ {\bQ}_\a, \bar{{\bQ}}_\b \} & = 2 \g^\m_{\a\b }( P_\m + Z_\m)+2i \e_{\a\b} Z~, \nonumber
\\[2mm]
\{ {\bQ}_\a, {\bQ}_\b \} & = i \g^\m_{\a\b}\z_{\m}~,\label{wallcharge}
\end{align}
where
\beq
Z= -\int d^2 x \; \e_{0\m\n} F^{\m\n}, \quad Z_\m= -\int d^2 x \; \e_{0\m\n} H^{\n}, \quad \z_\m= \int d^2 x \; \e_{0\m\n} \bar{Y}^{\n}. \label{brane charges}
\eeq
The first term, $Z$, is the usual central charge, and is carried by zero-branes (particles). The other two terms are carried by one-branes, which in three dimensions are domain walls. Thus for theories that support domain walls with these charges, there is a physical obstruction to removing these charges by an improvement transformation \cite{Dumitrescu:2011iu}. In particular, depending on which of these brane-currents can be improved to zero, we have different supersymmetry on the world-volume of the BPS-saturated domain walls. This is because the values of these brane-charges determine which combination of the supercharges are preserved in the domain-wall background. For instance, in theories which have an FZ-multiplet (for which $H^\nu=0$), domain walls carry the charge $\z_\m$ and hence by (\ref{wallcharge}) can only lead to non-chiral $(1,1)$  supersymmetry on the wall. On the other hand, for theories with an 
$\mc{R}$-multiplet (for which $Y^\nu=0$), domain walls carry the charge $Z_\m$ and hence by (\ref{wallcharge}) can only lead to a (2, 0) or (0, 2) chiral supersymmetric theory on the wall.  This will be important in section \ref{sec:U(1)_0}.
\subsection{Examples}
\subsubsection{Abelian gauge theory}\label{2.2.1}
For our purposes we focus on $U(1)$ gauge theories coupled to chiral superfields, including FI and Chern-Simons terms as well as a non-zero real mass. Even though when there is only one chiral superfield, the real mass can be absorbed by field-redefinitions, keeping it will make clear the generalization to case of multiple chiral fields. 
The Lagrangian \cite{Aharony:1997bx,Intriligator:2013lca} is given by
\beq
\mc{L}=\int d^4 \theta \; \left(  \frac{1}{4e^2} \Sigma^2 +\frac{k}{8\pi} \Sigma V -\bar{\Phi} e^{V+ V_b } \Phi + tV \right),
\label{N=2 Lagrangian}
\eeq
where $\Sigma= \frac{i}{2} \bar{{\bD}} \cdot {\bD} V$ is a real linear superfield that 
obeys $ \bar{{\bD}}^2 \Sigma={\bD}^2 \Sigma=0$, and $V_b$ is a constant superfield that generates the real 
mass $\Sigma_b\equiv \frac{i}{2} \bar{{\bD}} \cdot {\bD} V_b=2m$. The constants $k$ and $t$ are the Chern-Simons and the FI couplings respectively. The $\mcs$ superfield (in chiral representation; see appendix \ref{sym_A} for details) that satisfies the equations (\ref{N=2 S}) is given by
\beq
\mcs_{\a\b}= \bar{{\na}}_{(\a}  \left( e^{V+ V_b }  \bar{\Phi} 
\right) {\na}_{\b)}  \left( e^{V_b }  \Phi 
\right) - \frac{1}{2e^2}{\bD}_{(\a} \Sigma \; \bar{{\bD}}_{\b)} \Sigma ~,
\eeq
with ${\bY}_\a=0$ and 
\beq
{\bX}_\a= \frac{1}{8e^2} \bar{{\bD}}^2 {\bD}_\a \left(\Sigma^2\right)  + \frac{1}{2}  \bar{{\bD}}^2 {\bD}_\a \left(\bar{\Phi} e^{V+ V_b } \Phi \right) +4m i \bar{{\bD}}_\a \left( \bar{\Phi}e^{V+ V_b } \Phi  \right) -2it \bar{{\bD}}_\a \Sigma~.
\label{XisD Ab}\eeq
In the case of many chiral fields, one just needs to sum the two $\Phi$-dependent terms over all fields with the appropriate charges and masses. Since $\bY_\a=0$, the Abelian gauge theory has an $\mc{R}$-multiplet. Note that \eqref{XisD Ab} implies ${\bX}_\a = 8i \bar{{\bD}}_\a \mc{J}$, which means the brane current $H_\mu$ given by \eqref{H current} can be written as $H_\m = \pa_\m \mc{J}{\Big | }$, with $\mc{J}$ the linear multiplet
\beq
\mc{J} = \frac{i}{8}\bar{{\bD}} \cdot {\bD} \left( \frac{1}{4e^2}   \Sigma^2  + \bar{\Phi} e^{V + V_b} \Phi  \right) + \frac12m ~ \bar{\Phi} e^{V+V_b}  \Phi  -\frac14 t \Sigma~. \label{J ab}
\eeq
\subsubsection{\texorpdfstring{$SU(N)$}{SU(N)} gauge theory}
Here we derive the $\mcs$-multiplet for an $SU(N)$ gauge theory with a superpotential. The Lagrangian of such theory can be written in superspace as \cite{Gates:1991qn, Gaiotto:2007qi, Intriligator:2013lca},
\beq
\mc{L} = \int_{} d^4\theta \, \bigg( \frac{1}{2e^2}\Tr[\Sigma^2]
+\frac{k}{4\pi} \int_{0}^{1}dt \, \Tr[V \Sigma_t ]
- \bar{\Phi}_i e^{V} \Phi_i \bigg)
+ \left( \int_{} d^2\theta \, \mc{W}(\Phi) + \text{c.c.} \right), \label{Lagrangian}
\eeq
where  $\na$ is the gauge-covariant superderivative, $\Sigma = \frac{i}{2} \{ \bar{\na}^\a, \na_\a \} = \frac{i}{2} \bar{{\bD}}^\a (e^{-V} {\bD}_\a e^V)$ is the non-abelian scalar field strength written in the chiral representation (see appendix \ref{sym_A} for details), and is covariantly linear $\na^2 \Sigma = \bar{\na}^2 \Sigma = 0$, and  $\Sigma_t = \frac{i}{2} \bar{{\bD}}^\a (e^{-tV} {\bD}_\a e^{tV})$. The trace (in the fundamental representation) is normalized to $\Tr(T_\MM T_\NN)=\frac12\d_{\MM\NN}$.
The equations of motion are:
\bea
\bar{{\bD}}^2 \pa_i \mc{K} &= 4\pa_i \mc{W} ,\\[2mm]
\left( \bar{\na}\cdot\na \Sigma \right) &= \frac{i e^2 k}{2\pi} \Sigma - 2i e^2 \Phi_i \bar{\Phi}_i e^V + \left( \frac{2ie^2}{N} \bar{\Phi}_i e^{V} \Phi_i \right) \mathbbm{1}_{N \times N}, \label{eom for V}
\end{align}
where $\mc{K}=\sum_i \bar{\Phi}_i e^{V} \Phi_i$ is the K{\"a}hler potential and we have used $\Tr \left[ \Phi_i \bar{\Phi}_i e^{V} \right] = \bar{\Phi}_i e^{V} \Phi_i$, and also the fact that $\Sigma$ and its covariant derivatives are in the adjoint representation and hence traceless. After a detailed analysis using the equations of motions, a gauge invariant $\mcs$-multiplet satisfying equation \eqref{N=2 S} can be found as
\bea
\mcs_{\a\b} &= \left( \bar{\na}_{(\a} \bar{\Phi} \, e^V \right) \left( \na_{\b)} \Phi \right) - \frac{1}{ e^2} \Tr[ \left( \na_{(\a} \Sigma \right) \left( \, \bar{\na}_{\b)} \Sigma \right) ],\\
{\bX}_\a &= \frac{1}{4e^2} \bar{{\bD}}^2 {\bD}_\a \Tr \left[ \Sigma^2 \right] +\frac{1}{2} \bar{{\bD}}^2 {\bD}_\a \left( \bar{\Phi} e^{V} \Phi \right),\\
{\bY}_\a &=  4 {\bD}_\a \mc{W}.
\end{align}	
If the theory has a continuous $\mc R$-symmetry, i.e., if $\d \mc{W} \equiv \sum_{i}^{} R_i \Phi_i \pa_i \mc{W}=2\mc{W}$, then this supercurrent can be improved to an $\mc{R}$-multiplet by an improvement transformation given by $U = \sum_{i}^{} R_i \Phi_i \pa_i \mc{K}$ \cite{Dumitrescu:2011iu}, with $\mcs_{\a\b}\to \mc R_{\a\b}$:
\begin{align*}
\mc{R}_{\a\b} &= \left( \bar{\na}_{(\a} \bar{\Phi} \, e^V \right) \left( \na_{\b)} \Phi \right) - \frac{1}{ e^2} \Tr[ \left( \na_{(\a} \Sigma \right) \left( \, \bar{\na}_{\b)} \Sigma \right) ] + \frac{1}{2}\left[ {\bD}_{( \a}, \bar{{\bD}}_{\b)} \right]U~,\\
{\bX}_\a &= \frac{1}{4e^2} \bar{{\bD}}^2 {\bD}_\a \Tr \left[ \Sigma^2 \right] +\frac{1}{2} \bar{{\bD}}^2 {\bD}_\a \left( \bar{\Phi} e^{V} \Phi \right) - \bar{{\bD}}^2 {\bD}_\a U~,\\
{\bY}_\a &=  0~.
\end{align*}
We can also turn on real masses corresponding to flavor symmetries by introducing  constant background gauge superfields for the Cartan subalgebra of the flavor symmetry group. The real masses reside in the scalar component of the vector-multiplets corresponding to these background gauge fields, which we collectively call $V_b$. More precisely $\Sigma_b = 2M$, where $M$ is the real mass matrix. We find the following $\mc{R}$-multiplet
\bea
\mc{R}_{\a\b} &= \left( \bar{\na}_{(\a} \bar{\Phi} \, e^{V+V_b} \right) \left( e^{-V_b} \na_{\b)} e^{V_b} \Phi \right) - \frac{1}{ e^2} \Tr[ \left( \na_{(\a} \Sigma \right) \left( \, \bar{\na}_{\b)} \Sigma \right) ] +  \frac12[{\bD}_{(\a},\bar{{\bD}}_{\b)}] U~,\\
{\bX}_\a &= \bar{{\bD}}^2 {\bD}_\a \left( \frac{1}{4e^2} \Tr \left[ \Sigma^2 \right] +\frac{1}{2} \bar{\Phi} e^{V+V_b} \Phi - U \right)
+ 4i \bar{{\bD}}_\a \left( \bar{\Phi} e^{V+V_b} M \Phi \right) ,\label{XisD}\\
{\bY}_\a &=  0.
\end{align}
As in the abelian case,  we have ${\bX}_\a = 8i \bar{{\bD}}_\a \mc{J}$ and thus $H_\m = \pa_\m \mc{J}{\Big | }$, with $\mc{J}$ the linear multiplet
\beq
\mc{J} = \frac{i}{8}\bar{{\bD}} \cdot {\bD} \left( \frac{1}{2e^2} \Tr \left[ \Sigma^2 \right] + \bar{\Phi} e^{V + V_b} \Phi - 2U \right) + \frac12 \bar{\Phi} e^{V+V_b} M \Phi ~. \label{brane current}
\eeq
\section{\texorpdfstring{$\mcs$}{S}-multiplet for \texorpdfstring{$\mc{N}=1$}{N=1} theories \label{sec:3}}
\subsection{Defining equations}
As in the $\mc{N}=2$ theories, we can define a superfield for the stress-energy multiplet. Since the energy-momentum tensor is the highest-spin component of the multiplet, in this case the $\mcs$ multiplet is a spin $3/2$ superfield $\mcs_{\a\b,\g}=\theta^\d T_{\a\b,\g\d}+\dots$.We can determine the analog of (\ref{N=2 S}) for $\mc{N}=1$ theories by reduction: we define two copies of $\mc{N}=1$ superspace 
\beq
\{ D^{(I)}_\a, D^{(J)}_\b  \}= -i \d^{IJ} \pa_{\a\b} ~.
\eeq
The $\mcs$-multiplet for $\mc{N}=1$ theories can be defined as
\beq
\tilde{\mcs}_{\a\b,\g}= D_\g^{(2)}\mcs_{\a\b} \;|_{\theta^{(2)}=0}~;
\eeq
after some algebra we find the following equations (where we drop the superscript 
$D^{(1)}\to D$ since everything now refers to the first copy of the $\mc{N}=1$ superspace)
\bea
 D^\b \tilde{\mcs}_{\a\b,\g} &= D_{(\a} \tilde{X}_{\g)} + \e_{\a\g} D^\b \tilde{Y}_\b~, 
\label{n1s}\\[1mm]
 \tilde{\mcs}_{\a\b}\;^\b & =-2\tilde{Y}_\a ~,\\
 D^\b D_\a \tilde{X}_\b & =0~, \quad D^\b D_\a \tilde{Y}_\b=0~. \label{S eqn}
\end{align}
where $ \tilde{X}_\a$ and $ \tilde{Y}_\a$ are certain combinations of their $\mc{N}=2$ counterparts. Note (\ref{S eqn}) implies that $\tilde{X}_\a,\tilde Y_\a$ can be written (locally) in terms of potentials $\tilde X,\tilde Y$:
\beq
\tilde{X}_\a=D_a \tilde{X}, \quad \tilde{Y}_\a=D_a \tilde{Y}~, 
\eeq
Equations (\ref{S eqn}) are invariant under two transformations, which we parametrize by independent superfields $V$ and $U$:
\bea
 \d \tilde{X}  &=  V +3  D^2 U~, \nonumber\\
 \d \tilde{Y} &= -3  V - D^2 U~, \nonumber \\
 \d \tilde{\mcs}_{\a\b,\g} &= 2 \e_{\g (\a} D_{\b)} V -2i \pa_{\g(\a} D_{\b)} U~. \label{redef}
\end{align}
The transformation corresponding to $U$ gives rise to improvement transformations. The $V$ transformation can be used to set either $\tilde X$ or $\tilde Y$ equal to zero. Two convenient combinations invariant under $V$ transformations are
\bea
 X & = 3 \tilde{X} + \tilde{Y}~, \\
\mcs_{\a\b,\g} & = \tilde{\mcs}_{\a\b,\g}+\e_{\g (\a} D_{\b)} (\tilde{X}+\tilde{Y})~.
\end{align}
Under $U$ they transform as
\bea\label{simpr}
\d \mcs_{\a\b,\g} & = 4 D_{(\a} D_{|\g|}D_{\b)} U ~, \\[2mm]
\d X &= 8 D^2 U~.
\end{align}
Using the identity (\ref{pDDD}) we see that the divergence of the right-hand side of (\ref{simpr}) is zero, verifying that this is indeed an improvement transformation. In terms of these superfields, equations (\ref{n1s}-\ref{S eqn}) combine into
\bea
\mcs_{\a\b}\: ^\b & = D_\a X \nonumber~, \\[2mm]
D^\b \mcs_{\a\b,\g} & = D_\g D_\a X ~~\Leftrightarrow~~ D^\g \mcs_{\a\b,\g}=0~.
 \label{D2 on phi}
\end{align}

We have derived the defining equations \ref{D2 on phi} by reducing their known $\mc{N}=2$ counterparts. In the appendix \ref{sec:sugra} we rederive them by considering $\mc{N}=1$ linearized supergravity. This shows that this is indeed the most general $\mcs$ multiplet we can write down. 

As before, we can identify the components that correspond to the supercurrent and the energy-momentum tensor. From (\ref{D2 on phi}), the supercurrent can be identified as the lowest component of $\mcs_{\a\b,\g}$, while the energy momentum tensor as the lowest component of
\beq
T_{\a\b,\g\d}= -\frac{1}{4} \left(D_{(\d} \mcs_{|\a\b| \g)}+D_{(\a} \mcs_{|\g\d| \b)}\right)~.
\eeq
Note that the trace  the energy momentum tensor is $T \propto D^\a \mcs_{\a \b}\;^\b=D^2 X$. Hence, if $X$ can be improved to zero then the energy-momentum tensor is traceless implying scale invariance. As in the $\mc{N}=2$ case, there is also an additional tensor given by
\beq
H_{\a\g}= -\frac{i}{8} \pa_{\a\g}X~.
\eeq
This is a total derivative and therefore its curl is zero, implying that
\beq
H_{\a\b,\g\d}:=\e_{\a\g} H_{\b\d}+\e_{\b\g}H_{\a\d}+\e_{\a\d}H_{\b\g}+\e_{\b\d}H_{\a\g}~,
\eeq
is a conserved brane current:
\beq
\pa^{\a\b}H_{\a\b,\g\d}=0=\pa^{\g\d}H_{\a\b,\g\d}~.
\eeq

Having identified all the essential components of the $\mcs$-multiplet, we can calculate the anticommutator of a supercharge with a supercurrent:
\beq\label{n1brane}
\{ Q_\d, \mcs_{\a\b,\g} {\big|} \}= D_\d  \;\mcs_{\a\b,\g}{\Big | }= - \left(T_{\a\b,\g\d}+ H_{\a\b,\g\d}\right){\Big | }~.
\eeq
This shows that indeed $H_{\a\b,\g\d}$ is a brane current.  In vector notation, the energy-momentum tensor are the lowest component of the superfield
\bea
T_{\m\n} &= -\frac{1}{8}\left( (\g_\n)^{\g \d} D_{\d} \mcs_{\m \g} +(\g_\m)^{\g \d} D_{\d} \mcs_{\n \g} 
\right) ~,
\end{align}
and for the dual of the additional conserved brane current,
\beq
H_{\m}= -\frac{i}{8}\pa_{\m}X ~.
\eeq
Then (\ref{n1brane}) can be expressed as
\beq
\{ Q_\d, \mcs_\m{}^\g {\big|} \}=   2(\g^\n)_\d{}^\g \left( T_{\m\n} + \e_{\m\n\r} H^\r \right){\Big | }~. \label{commutator}
\eeq
When integrated over the whole space, the first terms on the right-hand side gives the usual momentum, while the second term is a brane charge
\beq
\{ Q_\a, Q_\b \}=  2\g_{\a\b}^\m (  P_\m + Z_\m) \label{extented algebra}~,
\eeq
where $Z_\m$ is a domain-wall charge
\beq
Z_\m=-\int d^2 x \; \e_{0\m \n} H^\n ~.
\eeq 
\subsection{Examples}
\subsubsection{Wess-Zumino \label{sec:WZ}}
Consider the Wess-Zumino model with Lagrangian
\beq
L=\int d^2 \theta \left( \frac{1}{2} D^\a \phi D_\a \phi + i \mc{W}(\phi) \right) \label{WZ Lagrangian}~.
\eeq
The equations of motion are
\beq
D^2\phi= i \mc{W}^\prime~.
\eeq
It can be verified that
\beq
\tilde{\mcs}_{\a\b,\g}= -2(D_\b D_\g \phi )D_\a \phi -  2(D_\a D_\g \phi) D_\b \phi~,
\eeq
satisfies equation (\ref{S eqn}) with
\beq
\tilde{X}_\a=  -D_\a \left( \frac{1}{2} (D\phi)^2 + i \mc{W} \right), \quad \tilde{Y}_\a=  D_\a \left( \frac{1}{2} (D\phi)^2 - i \mc{W} \right)~ .
\eeq
The brane current is therefore
\beq
H_{\a\b}=i \pa_{\a\b} \left( \frac{1}{8} (D\phi)^2 + \frac{1}{2} i \mc{W} \right){\Big | }= i  \pa_{\a\b} \left( \frac{1}{8} \psi^\a \psi_\a +\frac12 i \mc{W} \right)~. \label{H WZ}
\eeq
An important observation is that this is a total derivative. If we put the theory in a finite volume, and integrate (\ref{H WZ}) over all space to arrive at (\ref{extented algebra}), assuming that all fermions are zero in the vacuum, the first term in (\ref{H WZ}) vanishes. The second term, however, will contribute 
\beq
Z_\m = -\int\!\! d^2 x\, \e_{0\m\n}H^{\n} = \frac12\int\!\! d^2 x \,\e_{0\m\n} \pa^{\n} \mc{W}~.
\eeq
Using Lorentz symmetry, we can always consider a domain wall normal to the $x^1\equiv x$ direction. In this case $Z_0$ and $Z_1$ are zero while $Z_2$ is
\beq
Z_2 = \frac12 \int\!\!dy\, \Delta \mc{W} \label{Z1}~~,~~~~\Delta \mc{W}\equiv \mc W\,\Big|^{x=+\infty}_{x=-\infty}~.
\eeq
\subsubsection{Abelian gauge theory}
Although in the rest of the paper we don't study $\mc{N}=1$ gauge theories, we present here the $\mc{N}=1$-multiplet for a $U(1)$ gauge theory as another example. Consider the Langrangian 
\beq
L= \frac{1}{g^2}\int d^2 \theta  \left(W^\a W_\a + \frac{k}{2} \; \Gamma^\a W_\a  \right)~.
\eeq
The field strength $W_\a$ is defined by
\beq
W_\a=\frac{1}{2} D^\b D_\a \Gamma_\b~,
\eeq
and satisfies 
\beq
D^\a W_\a=0 \Leftrightarrow D_\a W_\b=D_\b W_\a~.
\eeq
The equations of motion are
\[ i\pa_\a{}^\d W_\d + 2 k \; W_\a=0~. \]
The $\mcs$-superfield is equal to
\beq
\tilde{\mcs}_{\a\b,\g}= W_\a D_\b W_\g+ W_\b D_\a W_\g ~. 
\eeq
with
\beq
\tilde{X}_\a=- \tilde{Y}_\a= \frac{1}{4}  D_\a (W^\b W_\b)~.
\eeq
The brane current is just the lowest component of
\beq
H_{\a \b}= -\frac{i}{16}  \pa_{\a\b} (W^\g W_\g)~;
\eeq
as usual, the lowest component of  $W_\a $ is the gaugino. Just as in four dimensions, 
gaugino condensation gives rise to a brane charge \cite{Dvali:1996xe}.
\section{Domain walls \label{sec:DM}}
In any theory with degenerate vacua, we can consider classical configurations that interpolate between two different degenerate vacua at plus and minus infinity and minimize the energy; these are domain walls. These objects carry the topological charges that were discussed above and therefore are topologically stable configurations. As we will see in the next sections, domain walls provide a tool to test various proposed dualities by considering such configurations on both sides of the duality. Before we move on, we review some well-known basic knowledge about BPS domain walls in supersymmetric theories.
\subsection{\texorpdfstring{$\mc{N}=1$}{N=1} theories}
Consider the anticommutator of the supercharges in the $\mc{N}=1$ superalgebra
\beq
\{ Q_\a, Q_\b \}=  2\g_{\a\b}^\m (  P_\m+ Z_\m)~.
\eeq
Since the brane charge $Z_\m$ is conserved on a domain wall with normal vector $n^\m$, it has to satisfy
\beq
n^{\m} Z_{\m}=0~,
\eeq
Going to the rest frame of the wall we have 
\beq
\{ Q_\a, Q_\b \}= 2 \d_{\a\b} m+ 2 \g_{\a\b}^\m Z_{\m}~, \label{N=1 Algebra}
\eeq
where $m$ is the total mass of the wall. Because the left-hand side of \eqref{N=1 Algebra} is a positive definite matrix, we get the following BPS bound
\beq
m \geq \sqrt{{Z_1}^2 + {Z_2}^2}~,
\eeq
since $Z_0=0$ because the wall is static. These equations makes sense in a finite volume; to take the infinite volume limit, we replace the mass and charges with energy and charge densities on the wall. 

Taking the coordinates of the three-dimensional space to be $x^\m=(t,x,y) $, we can choose the wall to be along one spatial direction, say $y$; since we are in its rest frame, the normal vector is $n^\m=(0,1,0)$ and consequently $Z_1=0$. Let us now assume that this bound is saturated
\beq
m =|Z_2|,
\eeq
The (\ref{N=1 Algebra}) becomes
\beq
\{ Q_\pm,Q_\pm \}=2(|Z_2| \pm Z_2)~~,~ \quad \{ Q_+,Q_-\}=0~.
\eeq
Hence, we see that when $Z_2$ is negative we have a domain walls with $Q_+$ unbroken leading to $(1,0)$ supersymmetry, and when $Z_2$ is positive anti-domain walls with $Q_-$ unbroken and $(0,1)$ supersymmetry.
\subsection{\texorpdfstring{$\mc{N}=2$}{N=2} theories \label{sec:N=2 DW}} 
As mentioned at the end of subsection \ref{sec:def for N=2}, depending on the values of brane-charges we have different supersymmetry for the domain-wall theory. Here we discuss theories with an $\mc R$-symmetry, which allows us to set ${\bY}_\a=0$ by an improvement transformation. Examples of such theories are studied in the next section.

In these theories, the $\mc{S}$-multiplet can be improved to an $\mc{R}$-multiplet and therefore the domain wall is only charged under the $Z_\mu$ brane-charge. Thus, the algebra (\ref{wallcharge}) becomes
\beq
\{ {\bQ}_\a, \bar{{\bQ}}_\b \}  = 2 \g^\m_{\a\b }( P_\m + Z_\m)~~,~~~~
\{ {\bQ}_\a, {{\bQ}}_\b \}  =0~.
\eeq
As in the $\mc N=1$ case discussed above,  when $Z_2$ is negative, a domain wall along the $y$ direction preserves supercharges of one chirality--in this case ${\bQ}_+$ and  $\bar{{\bQ}}_+$. This leads to $(2,0)$ supersymmetry on the domain wall, while when $Z_2$ is positive we have $(0,2)$ supersymmetry. 
Similarly, for theories with an FZ-multiplet where ${\bX}_\a=0$ and ${\bY}_\a \neq 0$, domain walls are charged only under $\z_\m$ and the unbroken supersymmetry is $(1, 1)$. Thus, for theories with an $\mc{R}$-multiplet or FZ-multiplet the domain walls are half-BPS.
\section{Supersymmetry enhancement from \texorpdfstring{$\mc{N}=1$}{N=1} to\texorpdfstring{ $\mc{N}=2$}{N=2} \label{sec:5}} 
Recently, an interesting IR duality was proposed between an $\mc{N}=1$ Wess-Zumino model with an $SU(3)$ flavor symmetry and an $\mc{N}=2$ abelian gauge theory \cite{Gaiotto:2018yjh,Benini:2018bhk}. This duality implies that the WZ model has enhanced supersymmetry in the IR, while in the gauge theory, the flavor symmetry is enhanced to $SU(3)$. In \cite{Gaiotto:2018yjh}, it was shown that after deforming both sides, the phase diagrams matches exactly.
Following these two papers, a third dual model was proposed with manifest $\mc{N}=2$ supersymmetry and $SU(3)$ flavor symmetry but with no time-reversal symmetry \cite{Fazzi:2018rkr}. In this section we consider domain walls in all three models and show that, as expected from the duality, on the wall we get the same two-dimensional effective theory in all three models. More specifically, in all three cases we show that the domain walls have a $(2,0)$ theory with target space $ {\bR}\times S^1$, while anti-domain walls give a $(0,2)$ theory with the same target space. 
\subsection{\texorpdfstring{$SU(3)$}{SU(3)} WZ model \label{sec:DW in WZ}}
Consider the $\mc{N}=1$ Wess-Zumino model with an $SU(3)$ global symmetry with superpotential
\beq
\mc{W}=\tr( \frac{2}{3}\phi^3 + M \phi)
\eeq
studied in \cite{Gaiotto:2018yjh}.
Both $\phi$ and $M$ transform in the adjoint representation of the $SU(3)$. The phases of the model with $M=m_3 T_3+ m_8 T_8$ where $T_I$ are $SU(3)$ matrices normalized as $\tr T_I T_J=\frac{1}{2} \d_{IJ}$, were analyzed in \cite{Gaiotto:2018yjh}. It was showed that for $m_3 \neq 0, \pm \sqrt{3}m_8$ the theory has two discrete vacua with the $SU(3)$ symmetry broken down to the $U(1)\times U(1)$ generated by $T_3$ and $T_8$. 

We are interested in domain walls interpolating between these two vacua. For simplicity we focus on the region  
$ 0< m_3 < \sqrt{3} m_8$. The two vacuum solutions of $\pa_I \mc{W}=0 $ are
\beq
\phi_{3,cl}^\pm=\pm \frac{ 3^\frac{1}{4} \left(m_8-\mu\right) \sqrt{\mu+m_8}}{\sqrt{2} m_3}~, \quad
\phi_{8,cl}^\pm=\pm \frac{3^\frac{1}{4} \sqrt{\mu+m_8}}{\sqrt{2}}~.
\eeq
with all the other $\phi's$ set to zero and $\mu=\sqrt{m_3^2+m_8^2}$. Plugging these values at the superpotential we get that $\mc{W}^+>\mc{W}^-$. Hence, a domain wall $(Z_2<0)$ interpolates between the $+$ vacuum at $-\infty$ and the $-$ vacuum at $+\infty$, while an anti-domain wall interpolates between the $-$ vacuum at $-\infty$ and the $+$ vacuum at $+\infty$. Consider the first case
\beq
\phi_3(x \rightarrow \pm \infty) = \phi_{3,cl}^\mp, \quad \phi_8(x \rightarrow \pm \infty) = \phi_{8,cl}^\mp~. \label{boundary cond} 
\eeq
Since the brane charge in (\ref{Z1}) is negative, the unbroken bulk supercharge is $Q_+$. On the domain-wall background all fermions are zero, and demanding that their transformations under the unbroken supercharge are also zero, we obtain the BPS equations 
\beq
\pa_x{\phi}_I= -\pa_I \mc{W}~.\label{WZ BPS equations }
\eeq
The first observation is that there are no solutions to (\ref{WZ BPS equations }) with the boundary conditions (\ref{boundary cond}) such that only $\phi_3$ and $\phi_8$ are activated. To see this, consider the BPS equations with all the other $\phi's$ set to zero
\beq
\pa_x{\phi}_3=-\frac{1}{2}m_3- \frac{1}{\sqrt{3}} \phi_3\phi_8, \quad 
\pa_x{\phi}_8=-\frac{1}{2}m_8-\frac{1}{2\sqrt{3}}(\phi_3^2-\phi_8^2)~.
\eeq
This set of equation has an ``integral of motion", which is cubic in the fields
\beq
G=-\frac1{2\sqrt3}\left(\sqrt{3} (m_3 \phi_8-m_8 \phi_3)+ \phi_3 \phi_8^2 -\frac{1}{3} \phi_3^3
\right).
\eeq
Since the value of $G$ is not the same for the two vacua, there is no solution with such boundary conditions. This implies that some of the other $\phi$'s must be activated and therefore at least one of the two $U(1)$'s is broken by the classical domain-wall solution. 

We have not solved equations (\ref{WZ BPS equations }) analytically, and so we rely on other methods to understand the qualitative properties of the solutions. We can do a stability analysis and extract all the necessary features that we need for the subsequent discussion. As a first step we use Morse theory \cite{Witten:1982im} (for a review see \cite{Hori:2003ic, Gaiotto:2015aoa}) to determine the dimension of the moduli space of solutions. The Morse function in this case is the superpotential itself. A straightforward calculation shows that the Morse index, i.e., the number of negative eigenvalues, at the two vacua is five and three respectively. That means that the space of solutions has dimension 2 (the difference of the Morse index at the two vacua), and hence the domain wall should have two bosonic moduli. However, we can do more, and study the topology of this space.

It is clear that one of the moduli is simply a translation in the $y$ direction and hence has the topology of $\mathbb{R}$.
The vacuum solutions that the domain wall tends to at $x\to\pm\infty$ preserve two $U(1)$'s, but as we just saw, the full domain-wall background must involve more than just $\phi_{3,8}$, and hence at least one $U(1)$ is broken; both $U(1)$'s cannot be broken, as that would provide too many moduli. Hence one combination of the $U(1)$'s is unbroken, and the phase corresponding to the broken $U(1)$ is the second modulus; its topology is $S^1$, which implies the moduli space is 
\beq
\mc{M}={\bR} \times S^1~.
\eeq
As usual, for each bosonic modulus we have a fermionic zero mode. Consequently, upon quantization, the theory on the wall will be described by two massless scalar fields and two massless fermion fields, and therefore, on the wall we will two supersymmetries. In the bulk, there are two supercharges but one of them is broken on the domain wall and we see that there is an emergent supersymmetry. To decide if the two-dimensional theory is $(1,1)$ or $(2,0)$ we use the fact that the Morse index for the full Hessian 
$\pa_I\pa_J\mc{W} $ is 2, and therefore the index of the Dirac operator is also 2, which implies $(2,0)$ supersymmetry\footnote{This argument only holds for real fields, and hence $\mc N=2$ Wess-Zumino models can have domain walls with $(1,1)$ supersysmmetry.}.
Note that states charged under the unbroken $U(1)$ are all massive.

\subsection{\texorpdfstring{$U(1)_0$}{U(1)} gauge theory \label{sec:U(1)_0}}
In this section we study the $\mc N=2$ $U(1)$ gauge theory considered in \cite{Gaiotto:2018yjh} with two chiral fields of charge one and real masses $m_{1,2}=\pm m$, and $\Sigma_b|=2m$ (see discussion above in section \ref{2.2.1}):
\beq
\mc{L}=\int d^4 \theta \; \left(\frac{1}{4e^2} \Sigma^2 + tV -\bar{\Phi}_1 e^{V+V_b } \Phi_1-\bar{\Phi}_2 e^{V- V_b } \Phi_2 \right),
\label{5:N=2 Lagrangian}
\eeq
The theory has two vacua with
\begin{enumerate}
\item   $\s=m, \quad \vp_1=0, \quad |\vp_2|^2=t ~$,
\item  $\s=-m, \quad \vp_2 =0, \quad |\vp_1|^2=t ~$.
\end{enumerate}
Following the discussion in the subsection \ref{sec:N=2 DW}, the domain walls are charged only under the H-current which can be calculated from (\ref{J ab}) to be
\beq
H_\m= \pa_\m \left( \sum_i (\s+m_i) |\vp_i|^2- t \s \right)~. \label{Hm}
\eeq
Assuming that at $x\rightarrow - \infty$ the theory is at vacuum (1) and at $x\rightarrow + \infty$ the theory is at vacuum (2) we find that brane charge density is
\beq
Z_2= -2l_y mt ~,~~Z_0=Z_1=0~,
\eeq
where $l_y$ is the size of the compactified $y$ direction (cf.~equation \ref{Z1}). Thus for negative $Z_2$ there are two unbroken supercharges, namely $\bQ_+$ and $\bar \bQ_+$, see section \ref{sec:N=2 DW}. Demanding that the variation of the fermion fields for these two combinations is zero, we obtain the following BPS equations
\bea
\left\{\bar \bQ_+, e^{-q_i V_b} {\nabla}_\a e^{q_i V_b}  \Phi_i\right\} & =0 \quad 
\Rightarrow  \quad \bar{\nabla}_+\; e^{-q_iV_b} {\nabla}_\a e^{q_iV_b}  \Phi_i =0 ~,\nonumber\\
\left\{\bar \bQ_+, {\bD}_\a \Sigma \right\} & =0 \quad \Rightarrow  
\quad \bar{{\bD}}_+{\bD}_\a \Sigma=0 \nonumber~.
\label{DW equations}
\end{align} 
After eliminating the auxiliary fields, the lowest components of these equations can be rewritten as
\bea
\mc{D}_x \vp_i=( \s+m_i) \vp_i,&  \quad \mc{D}_t \vp_i-\mc{D}_y \vp_i=0~, \\
\quad \pa_x \s= e^2(\sum_i|\vp_i|^2-t), & \quad \pa_t \s -\pa_y \s=0~,
\label{generalized BPS}
\end{align}
along with $F_{\m\n}=0$. Letting the fields $\vp_i$ and $\s$ depend only on the $x$ coordinate these equations simplify to \cite{Lambert:1999ix,Shifman:2002jm,Tong:2005un}
\beq
(\pa_x- iA_x) \vp_i=( \s+m_i) \vp_i, \quad \pa_x \s= e^2(\sum_i|\vp_i|^2-t), 
\quad  A_y=A_t, \quad F_{\m\n}=0~ \label{N=2 BPS eq}.
\eeq
Since $F_{\mu\nu}=0$ and our spacetime is contractable, we can choose $A_\mu=0$.
Using equations (\ref{DW equations}), it is easy to show that the $F$-current in (\ref{F current}) vanishes in the domain-wall background; since the $F$-current gives rise to a central charge rather than a brane-charge, this is expected.

To study the solutions of (\ref{N=2 BPS eq}) we rewrite1 them as
\beq
e^2 \frac{\pa h}{\pa \s}= \pa_x \s~, \qquad
\frac12\frac{\pa h}{\pa |\vp_i|} = \pa_x |\vp_i|~,
\eeq
where
\beq
h(\s,\abs{\vp_1},\abs{\vp_2})=(\s+m)\abs{\vp_1}^2+(\s-m)\abs{\vp_2}^2-t\s~.
\eeq
This is just the bottom component of $\mc{J}$ in \eqref{J ab}. Up to a rescaling in the fields, the BPS equations can be written as a gradient flow equation of the Morse function $h$. As in the WZ model, we use Morse theory to analyze the domain-wall solutions. Calculating the Morse index at the two critical points and using the same reasoning as in the WZ case, we find that there is one modulus that describes the solutions of these equations, namely the position $x_0$. Note that rescaling the fields does not change the signature of the Hessian of the Morse function and therefore we can directly calculate the index using $h$. There are two more free parameters: the phases of $\vp_1$ and $\vp_2$. However, only the phase difference is physical, since the sum can be set to zero by a gauge transformation. 
This physical $U(1)$ is broken by the domain-wall solution, and hence, as in the WZ case, the moduli space of solutions is 
\beq
\mc{M}={\bR} \times S^1~.
\eeq
Since the target space is flat it supports maximal supersymmetry. Upon quantization these two moduli will become massless fields living on the domain wall along with two fermionic zero modes. In this case the two supersymmetries on the wall are generated by the two unbroken bulk supercharges $\bQ_+$ and $\bar \bQ_+$ giving rise to a $(2,0)$ theory. 
We now argue that in addition to the broken $U(1)$ flavor symmetry on the wall, there is an unbroken $U(1)$ whose charged excitations are massive fluctuations of the wall.

The bulk theory has, in addition to the gauged $U(1)$,  a $U(1)_F\times U(1)_T$ symmetry. The $U(1)_F$ flavor symmetry is broken by the wall discussed above; the $U(1)_T$ is topological and is generated by 
$J^\m_T\propto\e^{\m\n\r}F_{\n\r}$. We review what happens to the gauge field on the wall \cite{Shifman:2002jm}. The common phase of $\vp_1,\vp_2$ can be gauged away; then we quantize the theory on the wall by allowing the two moduli to depend on the two coordinates along the wall 
\beq
\vp_1= e^{i \frac{\l}{2}} \chi_1(x-x_0)~, \quad \vp_2= e^{-i \frac{\l}{2}} \chi_2(x-x_0)~,
\eeq
where $\l=\l(t,y)$ and $x_0=x_0(t,y)$ and $\l$ has the gauge invariant definition
\beq
\l=arg(\vp_1)-arg(\vp_2)~.
\eeq
The bulk Lagrangian contains the terms 
\bea
\mc{L} & = -| (\pa_\m-iA_\m) \vp_1|^2 -| (\pa_\m-iA_\m) \vp_2|^2 +\dots \\
&=  -(|\chi_1'|^2+|\chi_2'|^2) (\pa_i x_0)^2 - |\chi_1|^2 | \frac{1}{2} \pa_i \l - A_i|^2-  |\chi_2|^2 | \frac{1}{2} \pa_i \l + A_i|^2+ \dots
\end{align}
To minimize the action, the gauge field should be 
\beq\label{523}
A_j (t,x,y)= \frac{1}{2} f(x) \pa_j \l (t,y)~,~\hbox{with} ~f(x \rightarrow \pm \infty)=\pm 1~.
\eeq
 To determine $f(x)$, we have to write the whole action including the gauge-field kinetic term as a $x$-integral times the two-dimensional $(t,y)$. Extremizing, we would find a second order differential equation for $f(x)$, but this is not essential for our purposes. However, what  is important is that (\ref{523}) implies the gauge field along the wall obeys Dirichlet boundary conditions $F_{ij}=0$. Following \cite{Gaiotto:2008sa}, on a Dirichlet boundary the $U(1)$ gauge symmetry in the bulk becomes a global symmetry. Since both $\l$ and $ x_0$ are invariant under the gauge $U(1)$, it follows that only massive modes on the wall are charged under this global $U(1)$.

In conclusion the theory on the wall is a $(2,0)$ theory with the two supersymmetries generate by the two unbroken bulk supercharges, with a $U(1)$ global symmetry that couples only to massive modes. This matches exactly what we found for the WZ model in section (\ref{sec:DW in WZ}).
\subsection{\texorpdfstring{$SU(3)_{\frac{5}{2}}$}{SU(3)} model}
In this section we study the deformations of the model studied in \cite{Fazzi:2018rkr}. The theory is an $\mc{N}=2$ non-abelian $SU(3)_{\frac{5}{2}}$ gauge theory with three chiral superfields in the fundamental representation with a superpotential
\beq
\mc{W} = \e^{ijk}\e^{\mm \nn \mathbf{p}} \Phi_{i\mm} \Phi_{j\nn} \Phi_{k \mathbf{p}}~.
\eeq
Here $\{\mm,\nn,\pp\}$ are fundamental color indices and $\{i,j,k\}$ are fundamental flavor indices. This model has a manifest $SU(3)$ flavor symmetry rotating the three chiral superfields. After adding mass deformations, we study the vacuum solutions of the model and show that the phase diagram matches those of the other models of figure \ref{fig:1}. Then, for generic masses, we analyze the effective theory on domain walls and show that we get the same effective theory as in sections \ref{sec:DW in WZ} and \ref{sec:U(1)_0},  providing another check for the duality depicted in figure \ref{fig:1}. This theory has an $\mc R$-symmetry with $R_i = \frac23$ for the superfields $\Phi_i$.
\subsubsection{Vacuum equations for general \texorpdfstring{$SU(N)_{k}$}{SU(N)} models}
Here we derive the supersymmetry-preserving vacuum equations including the one-loop correction for an $SU(N)_k$ gauge theory with $N_f$ flavors in the fundamental representation. We add both real masses and mass terms in the superpotential corresponding to the $N_f-1$ Cartan generators of the global $SU(N_f)$. 
The only relevant quantum corrections are effective Chern-Simons terms, which are one-loop exact. The potential for the scalar fields is a sum of squares; for supersymmetric vacua, all these must vanish. There are three type of terms:
\begin{enumerate}
\item \textit{F-terms:} These are associated with the superpotential and give: $\pa \mc{W}=0$. 
\item \textit{Mass-terms:} Mass terms of the matter fields which give: $m^{i\nn}_\text{eff} \, \vp_{i\nn}=0$
\item \textit{D-terms:} $D$-terms of the gauge fields; these are the focus of this section.
\end{enumerate}
We derive the $D$-terms in the quantum corrected Lagrangian by using the fact that the real mass deformations are the lowest component of the background-flavor superfield strength $\Sigma_b$, and the fact that the quantum corrections will produce effective mixed CS terms for the color and flavor gauge fields.

The gauge group is broken by the VEV of $\s$, the lowest component of the color field-strength 
$\Sigma$,  and the $D$-term equation for the broken generators will simply be:
\beq
\sum_{i}^{}\bar \vp_{\mm i} {(T^{\cancel\NN})^\mm}_\nn \vp^{i \nn} = 0, \quad \text{where $T^{\cancel\NN}$ is a broken generator}.
\eeq
For the {\em unbroken} generators, which generically we can choose to be in the Cartan subalgebra of $SU(N)$, 
the Lagrangian includes:
\beq
\mc{L} \supset \int_{}^{}d^4\theta\, \frac{k^{\MM \NN}_\text{eff}}{8\pi} \Sigma^\MM V^\NN + \frac{k^{\MM I}_\text{eff}}{8\pi} \Sigma^\MM V^I_b
\supset  \frac{k^{\MM \NN}_\text{eff}}{2\pi} D^\MM \s^\NN
+\frac{k^{\MM I}_\text{eff}}{2\pi} D^\MM m^I~.
\eeq
Here $V^\MM$ are color gauge superfields, $V_b^I$ are background flavor gauge superfields, 
$m^I$ the real masses associated with Cartan generators of the flavor symmetry group, and $\MM,\NN,...$ are adjoint color  and $I,J,...$ are adjoint flavor indices. The complete $D$-terms for the Cartan generators $H^\MM$ in the Lagrangian are
\beq
\mc{L}_D = \frac{1}{2e^2}D^\MM D^\MM + \frac{D^\MM}{2\pi} \Big( k^{\MM \NN}_\text{eff}\s^\NN+k^{\MM I}_\text{eff}m^I- 2\pi\sum_{i}^{} \bar\vp_{\mm i} (H^\MM)^\mm_{\nn} \vp^{i \nn } \Big),
\eeq
where we restrict the color adjoint indices $\MM, \NN$ to the Cartan generators of the gauge group. Thus we get the $D$-term vacuum equations for the Cartan generators:
\beq\label{5.2Deq}
2\pi\sum_{i}^{} \bar\vp_{\mm i} (H^\MM)^\mm_{\nn} \vp^{i \nn} = k^{\MM \NN}_\text{eff}\s^\NN+k^{\MM I}_\text{eff}m^I.
\eeq
To calculate the effective Chern-Simons levels $k_\text{eff}$, we introduce the following notation for the Cartan generators: 
\beq
\text{color:}\quad{(H^\MM)^\mm}_\nn = {\d^\mm}_\nn g^{\mm \MM}
~~,~\quad\quad\text{flavor:}\quad{(H^I)^i}_j = {\d^i}_j n^{i I}~.
\eeq
Then
\bea
k^{\MM \NN}_\text{eff} &= k\d^{\MM \NN}+\frac{1}{2}\sum_{i,\mm}^{} g^{\mm \MM} g^{\mm \NN} \text{sign}(m^i+{\s}^\mm), \label{shift1}\\
k^{\MM I}_\text{eff} &= \frac{1}{2}\sum_{i,\mm}^{} g^{\mm \MM} n^{i I} \text{sign}(m^i+\s^\mm), \label{shift2}
\end{align}
where $\s=\text{diag}({\s}^\mm)$ with ${\s}^\mm = g^{\mm \MM} \s^\MM$, and $m^i=n^{i I} m^I$ is the real mass for the $i$-th flavor. Note that $m^i+{\s}^\mm = m_\text{eff}^{i\mm}$ is the effective mass for the chiral superfield $\Phi_{i \mm}$.
Equations \eqref{5.2Deq}, \eqref{shift1} and \eqref{shift2} then give:
\beq
2\pi\sum_{i}^{} \bar\vp_{\mm i} (H^\MM)^\mm_{\nn} \vp^{i \nn} = k\s^\MM + \frac{1}{2}\sum_{i, \mm}^{} g^{\mm \MM} \abs{m^i + {\s}^\mm}~.
\eeq
\subsubsection{Vacuum solutions}\label{vacsols5.3.2}
For the $SU(3)$ model, we write the real masses as $m^1$, $m^2$, $m^3=-m^1-m^2$ and the adjoint scalars as $\s=\text{diag}(\s^1,\s^2,\s^3=-\s^1-\s^2)$. Then
the supersymmetric vacuum equations are:
\begin{align}
&\pa_{i\mm} \mc{W}=0 ~~\qquad \Rightarrow ~~\qquad \vp_{i\mm}=q^iv^\mm~, \label{1}\\[1mm]
&(m^i + \s^\mm) \vp_{i\mm} = 0~, \label{2}\\[1mm]
&\sum_{i=1}^{3}\left(2\pi\abs{\vp_{i\mm}}^2- \frac{1}{2}\abs{m^i+\s^\mm}\right)= k\s^\mm +\frac13 \sum_{i,\nn=1}^{3}\Big(
2\pi\abs{\vp_{i\nn}}^2 - \frac{1}{2}\abs{m^i+\s^\nn}\Big)~, \label{3}\\
&\sum_{i=1}^{3}\bar\vp_{i\mm}\vp_{i\nn} = 0\quad \text{for} \quad \mm\neq \nn~.\label{4}
\end{align}
Note that in \eqref{1}, we have used the fact that for an $N\times N$ matrix $\vp$, $\d \, \text{det}(\vp)=0$ iff $\text{rank}(\vp)<N-1$.

Using equation \eqref{1} and \eqref{4}, we see that only one $v^\mm$ can be non-zero and by a residual gauge transformation we can take it to be $v^3$, so that $\vp_{i\mm} = \d_{\mm}^3 q_i$. 
Then substituting $k=\frac{5}{2}$, we arrive at a simpler set of equations:
\begin{gather}
(m^i + \s^3) q_i = 0,\\
0 = \frac{5}{2}(\s^1-\s^2) +\frac{1}{2}\sum_{i=1}^{3}(\abs{m^i+\s^1}-\abs{m^i+\s^2}), \label{sigma12}\\
2\pi\abs{q}^2 = \frac{5}{2}(\s^3-\frac{\s^1+\s^2}{2}) +\frac{1}{2}\sum_{i=1}^{3}(\abs{m^i+\s^3}-\frac{\abs{m^i+\s^1}+\abs{m^i+\s^2}}{2}),
\end{gather}
Now by using \eqref{sigma12} and the triangle inequality we get:
\beq
\frac{5}{2}\abs{\s^1-\s^2} = \abs{\frac{1}{2}\sum_{i=1}^{3}(\abs{m^i+\s^1}-\abs{m^i+\s^2})} \le \frac{3}{2}\abs{\s^1-\s^2},
\eeq
which is only possible for $\s^1=\s^2\equiv\s_0$. Since $\s$ is in the adjoint representation of $SU(3)$, we have
$\s^3=-2\s_0$ , we find:
\begin{gather}
(m^i -2\s_0) q_i = 0,\\
2\pi\abs{q}^2 = -\frac{15}{2}\s_0 +\frac{1}{2}\sum_{i=1}^{3}\left(\abs{m^i-2\s_0}-\abs{m^i+\s_0}\right). \label{equ}
\end{gather}
We now analyze the solutions of these equations in the generic case where $m^i\ne m^j$. There could be two kind of solutions:
\begin{enumerate}
\item \label{I}  $q_1=q_2=q_3=\s_0=0$.
\item \label{II} For one particular $i$: $q_j \equiv q\d_{ij}$ and $\s_0=\frac12m^i$.
\end{enumerate}
In the case \ref{I}., after integrating out the massive matter we will have an $\mc{N}=2$ $SU(3)_{k_{\text{eff}}}$ theory in the IR, where (\ref{shift1}) implies $k_\text{eff}=\frac{5}{2}+\frac12\sum_{i}^{}\text{sign}(m^i)$. Since the masses are in the Cartan subalgebra of the flavor $SU(3)$,
either two real masses are positive and one negative or two are negative and one positive (by assumption none vanish). This means that we would get either an $SU(3)_3$ or $SU(3)_2$; the Witten index of these theories is calculated in \cite{Intriligator:2013lca}, and implies only the  $SU(3)_3$ case does not break supersymmetry dynamically.

We now analyze equations \eqref{equ} and look for the solutions of type \ref{II}. The triangle inequality implies\beq
\frac{1}{2}\sum_{i=1}^{3}\left(\abs{m^i-2\s_0}-\abs{m^i+\s_0}\right) \le \frac{9}{2}\abs{\s_0}~,
\eeq
so the sign of the right hand side of equation \eqref{equ} is determined by the sign of $\s_0$, and therefore solutions with one $q_i\ne 0$ exist if and only if $\s_0<0$, which means the corresponding $m^i<0$. So for each negative $m^i$ we have a solution of type \ref{II}. In this case, the $SU(3)$ gauge group will be broken to $SU(2)$. To calculate the effective CS level we have to look at the sign of the effective masses for matter fields charged under the $SU(2)$. For simplicity assume that $\s_0=m^3/2$:
\beq
k_\text{eff} =
\frac{5}{2}+\frac{\text{sign}(m^1+\s_0)+\text{sign}(m^2+\s_0)+\text{sign}(m^3+\s_0)}{2}, 
\eeq
but, $m^1+m^2=-m^3$ implies $(m^1+\s_0)+(m^2+\s_0)=-m^3+2\s_0=0$ and hence
\beq
\text{sign}(m^1+\s_0)+\text{sign}(m^2+\s_0)=0~,~~~\Rightarrow~~~k_\text{eff}=2~.
\eeq
Thus the low-energy theory will be an $\mc{N}=2$ $SU(2)_{2}$ theory, which is trivial in the IR. To see this, note that the Chern-Simons level makes the $\mc{N}=2$ vector multiplet massive, which implies there are no light excitations and we have a TQFT. But the Witten index of this theory can be computed \cite{Intriligator:2013lca}, and it is equal to 1, so the TQFT must be trivial and we have the trivial phase in the IR.  

Therefore, depending on the signature of the mass matrix we have different kinds of solutions. Now we want to obtain the explicit form of the solution in each case. Without loss of generality we assume $m^3<0$ and $m^2>m^1$. Based on the sign of $m^1$ there are two cases which we study separately.
\begin{enumerate}
\item $m^2>-m^1>0$: In this case $m^1$ and $m^3$ are negative and we have two $SU(2)_2$ solutions with:
\beq\label{m1,m3<0}
\s =\frac{m^{1}}{2}~~ \text{or}~~\s =\frac{m^{3}}{2}~~, ~~\abs{q}>0~.
\eeq
\item \label{m2>m1} $m^2>m^1>0$: In this case only $m^3$ is negative and we have an $SU(3)_3$ solution with $\s=0$ and $q=0$ and an $SU(2)_2$ solution:
\beq\label{m1,m2>0}
\s=\frac{m^{3}}{2}~,~~\abs{q}>0~.
\eeq
\end{enumerate}
As a further check of the proposed IR duality of this model to those studied in sections \ref{sec:DW in WZ} and \ref{sec:U(1)_0}, we could obtain the full phase diagram of this model and compare it the phase diagram of the previous two models that was obtained in \cite{Gaiotto:2018yjh}. For generic mass deformations we already derived that there are two degenerate vacua confirming results of \cite{Gaiotto:2018yjh}. There are special loci on the phase diagram where there are some unbroken $SU(2)$ global symmetries, and they could result in a moduli spaces of vacua. These special loci arise when two of the real masses coincide, i.e. $m_i=m_j$. Without loss of generality we focus on the case with $m_1=m_2=m$. If $m>0$, then the situation is exactly similar to case \ref{m2>m1} above and we are in the same phase with two degenerate vacua. However, for $m<0$, there is a ${\bC}P^1$ Higgs moduli because, even though $\abs{q}^2$ is fixed, both $q_{1,2}$ can be nonzero. Thus for $m_1=m_2<0$, the theory flows to $\mc{N}=2$ ${\bC}P^1$ sigma-model which matches with the other two descriptions \cite{Gaiotto:2018yjh}.

We have studied the vacua of the theory, and found that due to quantum corrections, some classical vacua break supersymmetry dynamically, and thus at the quantum level are not degenerate with the supersymmetry preserving vacua. In the next subsection, we will use this information to only look at walls that interpolate between true supersymmetric vacua. However, we have only looked at the quantum corrections to the Chern-Simons level, and since these are not continuous functions of the fields, we cannot use them to find the quantum corrections to the domain walls themselves; in fact, we do not need to, and it suffices to study the classical BPS equations with the correct boundary conditions.
\subsubsection{BPS Domain Walls}
In this section we study the classical dynamics on domain walls of the above non-abelian gauge theory.  
As in the previous sections, we can evaluate the brane current \eqref{brane current} and for any pair of vacua, we can calculate the brane charge and depending on its sign, see if that pair gives rise to a domain wall or anti-domain wall. Then, from the unbroken supercharges we can obtain the BPS equations and further analyze them. 

Because we have a continuous $\mc R$-symmetry, the only non-zero brane current is the $H$ current whose its bosonic part is given by \eqref{brane current} as
\beq
H_\mu = \pa_\m\! \left( \frac{1}{2e^2} \s^\MM D^\MM -  \frac{1}{6} \abs{\vp_{i\nn}}^2 m_\text{eff}^{i\nn}(\s)  +\frac12\abs{\vp_{i\nn}}^2 m^{i}  \right),
\eeq
only the last term will contribute to the integrated current--the brane charge--since the other terms vanish by the vacuum equations. Taking $x$ to be the normal direction to the domain wall ($Z_2<0$), we find the brane charge 
\beq
Z_2 = -\int_0^{l_y}\!\!dy\,T_w = -\frac{1}{2} l_y\abs{\vp_{i\nn}}^2 m^{i} \Biggr|_{x=-\infty}^{x=+\infty}~~,~~~Z_0=Z_1=0~.
\eeq 
As in section \ref{sec:U(1)_0}, the domain wall preserves the two supercharges ${\bQ}_+ $ and $\bar{{\bQ}}_+$. Requiring that supersymmetry variations of the fermions with respect to these supercharges vanish, we get the BPS equations
\begin{align}
\pa_{i\mm} \mc{W} &= 0~, ~\qquad \qquad\bar\vp_{i\mm} {(T^{\cancel\NN})^\mm}_\nn \vp^{i\nn} = 0~,  \label{BPS1_5.3}\\[1mm]
e^2\frac{\pa h}{\pa \s^{\MM}}&= \pa_x \s^{\MM}~, \qquad
\frac{\pa h}{\pa \bar\vp_{i\mm}} = (\pa_x - i{A}^\mm_x) \vp_{i\mm}~, \label{BPS for sq}
\end{align}
with Morse function
\beq
h(\vp,\s) = \abs{\vp_{i\nn}}^2 m_\text{eff}^{i\nn}(\s) - \frac{k}{4\pi}(\s^\MM)^2~, \label{morse function}
\eeq
where $\cancel\NN$ runs over the broken gauge generators and
$m_\text{eff}^{i\mm}(\s) = {m}^i+{\s}^\mm(x)$ is the field-dependent effective mass of $\Phi_{i\mm}$. 

As for the vacuum equations \eqref{1}, 
\eqref{BPS1_5.3} imply that the most general solution will have the form 
$\vp_{i\mm}=w_i v_\mm$, and we can always choose a gauge 
$\vp_{i\mm}=\d_{\mm,3}q_i$.
Because of the gauge fields in the BPS equations \eqref{BPS for sq}, these equations are not quite the same as the Morse flow of $h$. 

The BPS equations, \eqref{BPS for sq} and \eqref{BPS1_5.3}, decouple into two set of equations for the gauge invariant variables, $\s^\MM$ and $\abs{q_i}$ ($i=1,2,3$), and gauge dependent variables, $A_x^3$ and $\theta_i$, where $q_j = \abs{q_j} e^{i\theta_j}$. The equations for the gauge invariant variables are the usual Morse flow equations of the Morse function $h$ up to rescaling the fields, and can be analyzed by standard Morse theory; the equations for the gauge dependent variables are
\beq
A^3_x - \partial_x \theta_j = 0, \quad \text{for } j=1,2,3~. 
\eeq
The differences $\theta_1-\theta_2$ and $\theta_2-\theta_3$ are gauge-invariant, and give rise to two flavor
$U(1)$ symmetries, whereas the sum, $\theta_1+\theta_2+\theta_3$, can be absorbed into $A_x^3$ by a gauge transformation. For each broken flavor $U(1)$ symmetry, we get an extra $S^1$ modulus. As we will see, in the model we study in this section only one is broken and we have only a single $S^1$ modulus. We encounter more complicated situation in section \ref{sec:2 to 4}.

We first ignore these phases and study at the BPS equations for $\abs{q_i}$ and $\s$. Recall $\s = \text{diag}(\s_1, \s_2, -\s_1-\s_2)$; then the Morse function is
\beq
h(\s_1,\s_2,\abs{q_i}) = \abs{q_i}^2(m^i-\s_1-\s_2) - \frac{k}{4\pi} {\left(\frac{\s_1+\s_2}{2}\right)}^2 -\frac{k}{4\pi} {\left(\frac{\s_1-\s_2}{2}\right)}^2. 
\eeq
The BPS equation for $\s_1-\s_2$ decouples from the rest and by Morse theory it does not have a non-zero solution. Hence we may set $\s_1=\s_2\equiv \s_0$ and find the reduced Morse function
\beq
h(\s_0,\abs{q_i}) = \abs{q_i}^2(m^i-2\s_0) - \frac{k}{4\pi} \s_0^2. 
\eeq
We now find the critical points of this function and their Morse indicies. 
As we saw above, there are two cases depending on the signature the real masses.
\begin{enumerate}
\item When two masses are negative, and one positive, e.g.,  $m_1<m_3<0<m_2$, we found two acceptable vacua (\ref{m1,m3<0}):
\begin{enumerate}
\item
\bea
~&~\nonumber\\[-15mm]
\s_0 &= \frac{m_1}{2}~~,~~q_1 = \sqrt{-\frac{k}{4\pi}\frac{m_1}{2}}~~,~~q_2=q_3=0~,~ 
\text{which has}\nonumber\\
h &= -\frac{km_1^2}{16\pi}~~\text{and Morse index}~ \mu=1~.
\end{align}
\item
\bea 
~&~\nonumber\\[-15mm]
\s_0 &= \frac{m_3}{2}~~,~~q_3 = \sqrt{-\frac{k}{4\pi}\frac{m_3}{2}}~~,~~q_1=q_2=0~,~
\text{which has}\nonumber\\
h &= -\frac{km_3^2}{16\pi}~~\text{and Morse index}~ \mu=2~.
\end{align}
\end{enumerate}
Hence, Morse theory implies there is only $2-1=1$ modulus--the translational mode--for these variables. Also, only one of $q_i$'s is non-zero, so only one of the remaining $U(1)$ symmetries is broken and hence there is an $S^1$ modulus. Putting these two moduli together, we get ${\bR}\times S^1$ sigma model on the wall. 
\item When two masses are positive, and one negative, e.g.,  $m_1,m_2>0>m_3$, we found two acceptable vacua (\ref{m1,m2>0}):
\begin{enumerate}
\item 
\bea
~&~\nonumber\\[-16mm]
\s_0& = \frac{m_3}{2}~~,~~q_3 = \sqrt{-\frac{k}{4\pi}\frac{m_3}{2}}~~,~~q_1=q_2=0~,~
\text{which has}\nonumber\\
h &= -\frac{km_3^2}{16\pi}~~\text{and Morse index}~ \mu=1~.
\end{align}
\item
\bea
~&~\nonumber\\[-16mm]
\s_0 &= q_1=q_2=q_3=0~,~
\text{which has}\qquad\qquad\qquad\qquad\qquad \nonumber\\
h &=0~~\text{and Morse index}~ \mu=2~.
\end{align}
\end{enumerate}
Again, Morse theory implies we have just the translational zero mode,  and the $S^1$ modulus comes from the phase corresponding to the broken $U(1)$ symmetry, giving an ${\bR}\times S^1$ sigma model on the wall.
\end{enumerate}
Thus the low energy theory on the domain wall is again an ${\bR}\times S^1$ sigma model, which matches the conjectured dual descriptions. This serves as a non-trivial check of the conjectured dualities. 
In particular, the ${\bR}\times S^1$ sigma model has (2,0) supersymmetry.
\section{\texorpdfstring{$\mc{N}=1$}{N=1} \texorpdfstring{$SU(5)$}{SU(5)} WZ Model  \label{sec:6}}
In this section, we consider the $SU(5)$ generalization of the WZ model studied in section \ref{sec:DW in WZ}, and investigate the possibility of supersymmetry enhancement in the IR. We consider the mass deformed theory with the superpotential
\beq
\mc{W} = \Tr( \frac{1}{3}\Phi^3 + M \Phi),
\eeq
where $\Phi$ is a traceless Hermitian matrix of real superfields, and $M$ is the real traceless mass deformation matrix. As in the $SU(3)$ WZ model in section \ref{sec:DW in WZ}, the deformed phases have exact moduli spaces of vacua which can be computed by solving the vacuum equations $\pa \mc{W} = 0$,
\beq
\Phi^2 + M = \frac{1}{5}\Tr(\Phi^2 + M).
\eeq
Depending on the eigenvalues of matrix $M$, we have different phases; these are summarized in Table \ref{table:1}.
\begin{table}[ht]
\centering
\begin{tabular}{ |c|c|c| }
	\hline
	Hypersurfaces & $\mc{M}$ & Phases \\
	\hline \hline
	(5) & $pt$ & $\l_i=0$ \\ [6pt]	
	(4,1) & $\text{Gr}(2,4)$ & $(\l_{1,2,3,4}<\l_5)$ \\ [6pt]
	(4,1) & $2 \, {\bC}\text{P}^3+S^0$ & $(\l_1<\l_{2,3,4,5})$ \\ [6pt]
	(3,2) & $2 \, {\bC}\text{P}^2$ & $(\l_{1,2,3}<\l_{4,5})$ \\ [6pt]
	(3,2) & ${\bC}\text{P}^1 + S^0$ & $(\l_{1,2}<\l_{3,4,5})$ \\[6pt]
	(2,2,1) & ${\bC}\text{P}^1 \times {\bC}\text{P}^1 + S^0$ & $(\l_{1,2}<\l_{3,4}<\l_5)$ \\[6pt]
	(2,2,1) & $2 \, {\bC}\text{P}^1 + 2 \, {\bC}\text{P}^1 + S^0$ & $(\l_{1,2}<\l_3<\l_{4,5}) 
	\lor (\l_1<\l_{2,3}<\l_{4,5})$ \\[6pt]
	(3,1,1) & $2 \, {\bC}\text{P}^2 + S^0 + S^0$ & $\frac{\l_1+\l_2}{2}-\frac{\sqrt{5}}{6}\abs{\l_1-\l_2}<\l_{3,4,5}$ \\[6pt]
	(3,1,1) & $2 \, {\bC}\text{P}^2$ & $\frac{\l_1+\l_2}{2}-\frac{\sqrt{5}}{6}\abs{\l_1-\l_2}>\l_{3,4,5}$ \\[6pt]
	(2,1,1,1) & $2 \, {\bC}\text{P}^1 + S^0,2\,{\bC}\text{P}^1+3S^0$ & Not known \\ [6pt] 
	(1,1,1,1,1) & $3S^0,5S^0,...$ & Not known\\
	\hline
\end{tabular}
\caption{The structure of moduli spaces of vacua for the mass deformed $\mc{N}=1$ $SU(5)$ WZ Model. Each phase has been labeled by a sequence indicating the order of distinct eigenvalues of $M$.  Here $S^0$ means two isolated vacua related by time reversal. For the last two phases, the vacuum equations cannot be solved analytically and the results are presented from numerical analysis.}
\label{table:1}
\end{table}
All the phases are either trivial or have moduli spaces of vacua that are K{\"a}hler manifolds. Thus the low energy theory describing these theories are sigma models with K{\"a}hler target space and have supersymmetry enhancement from $\mc{N}=1$ to $\mc{N}=2$ in the IR: at low scales, the only relevant term is the sigma-model kinetic term, which has $\mc{N}=2$ supersymmetry.

In light of this observation, one might guess that the undeformed theory has $\mc{N}=2$ enhancement as well. We investigate this hypothesis by studying domain-wall solutions. We consider the mass deformation matrix, $M$, to be generic, i.e., breaking the $SU(5)$ global symmetry down to $U(1)^4$. By symmetry arguments \cite{Gaiotto:2018yjh}, the classical vacuum equations will be exact and will not receive quantum correction. To study the vacuum equations, without loss of generality, we chose $M=\text{diag}(m_1,...,m_5)$. Now by setting $\d \, \mc{W}=0$, we find the vacuum equations,
\beq
\Phi^2 + M = \frac{1}{5}\Tr(\Phi^2 + M).
\eeq
First, we try to solve for $x=\frac{1}{5}\Tr(\Phi^2 + M)$. Since $M$ is diagonal and has distinct eigenvalues, $\Phi$ has to be diagonal as well and will have the form,
\beq
\Phi = \text{diag}(\vp_i), \quad \text{where:} \quad \vp_i = \pm \sqrt{x - m_i}.
\eeq
But since we are only interested in Hermitian and traceless $\Phi$, we should have,
\beq
\sqrt{x - m_1} \pm \sqrt{x - m_2} \pm \sqrt{x - m_3} \pm \sqrt{x - m_4} \pm \sqrt{x - m_5} = 0, \label{equation for x}
\eeq
for any given choice of signs, as well as
\beq
x \ge \max_{1 \le i \le 5} \{m_i\}. \label{reality condition for x}
\eeq
Now if we take the product of the LHS of Eq. \eqref{equation for x}, we get a polynomial of degree 8 in $x$, whose roots will be the solutions we are looking for. We cannot find the roots analytically in terms of $m_i$'s, but numerically, the polynomial has either 3 or 5 real roots satisfying condition \eqref{reality condition for x}. In the generic case, each solution for $x$ gives exactly two solutions for $\Phi$ which are related by time reversal. Thus, we have either six or ten isolated supersymmetry vacua in the generic case. We now study the phase with ten vacua and consider domain walls interpolating between these vacua.

Naively, we might think that for each pair of vacua out of the ten vacua we have a stable BPS-saturated wall interpolating between them. But it is not always true. Though a BPS-saturated wall minimizes the energy locally, it might not be true globally, i.e., it might be more efficient to first interpolate to an intermediate vacuum and then go to the final vacuum, so a bound state of two BPS-saturated walls might have a lower energy. Thus, for a BPS-saturated wall to be stable, it is necessary to satisfy the following triangle inequality \cite{Chibisov:1997rc},
\beq
\abs{\Sigma_{ij}} \le \abs{\Sigma_{ik}} + \abs{\Sigma_{kj}}, \label{triangle inequality}
\eeq  
where $\Sigma_{ij}$ is the central charge of the BPS-saturated wall interpolating between vacuum $i$ and $j$.
In our case, we have a WZ model and the central charges are given by $\Sigma_{ij}= 2(W_i - W_j)$, where $W_i$ is the value of the superpotential at the $i$-th vacuum. According to Morse theory, a wall interpolating between vacuum $W_i$ and $W_j$ generically has $n_{ij}=\mu_{j}-\mu_i$ zero modes when $n_{ij}$ is positive, and there are no BPS-saturated walls at all when $n_{ij}\le0$. 
Thus, to find stable walls 
$\bW_{ij}$ between vacua $W_i<W_j$, the Morse index must be increasing: $\mu_i<\mu_j$ and there must not be any intermediate vacua $W_k$ satisfying both conditions $W_i<W_k<W_j$ and $\m_i<\m_k<\m_j$. Note that for a wall interpolating between vacuum $W_i$ and $W_j$, if there exists a vacuum $W_k$ with, $W_i < W_k< W_j$, we have equality in \ref{triangle inequality}.

We now apply these constraints to study the solutions of the BPS equations. There are ten vacua with values of the superpotentials $W_1 < ... < W_{10}$, and the Morse indices $\mu_1, ..., \mu_{10}$.  The theory on this wall defines a 2d sigma model with an $n_{ij}$-dimensional manifold as the target space. For this sigma model to have supersymmetry enhancement, it must have at least $(2,0)$ supersymmetry, which requires the manifold to be even-dimensional. Numerically we find even and odd Morse indices at the vacua, which is incompatible with supersymmetry enhancement.

More explicitly, the Morse indices that we find for the phase with ten vacua are,
\beq
(\mu_1,...,\mu_{10})=(8,10,12,14,15,9,10,12,14,16)~.
\eeq
Thus, there are four BPS-saturated walls that have exactly one modulus: ${\bW}_{1,6}$,  ${\bW}_{4,5}$, ${\bW}_{6,7}$, and ${\bW}_{5,10}$. This modulus corresponds to the broken translational symmetry, and so the theory on these walls will be the 2d $\mc{N}=(1,0)$ theory of a massless scalar and a left-handed fermion. The remaining eight stable walls have an two moduli: 
$\bW_{1,2},\bW_{2,3},\bW_{3,4},\bW_{7,8},\bW_{8,9},\bW_{9,10},\bW_{2,8},\bW_{3,9}$,
and give a 2d $\mc{N}=(2,0)$ theory on the wall. 

In the phase with six vacua, the Morse indices are all even:
\beq
(\mu_1,...,\mu_{6})=(8,10,12,12,14,16)~.
\eeq
Hence, in this phase, there is supersymmetry enhancement on the walls, and in the phase with ten isolated vacua there need not be. 

In conclusion, we find for the generic relevant deformations of the $SU(5)$ WZ model considered here, there are domain walls with {\em no} supersymmetry enhancement. For the undeformed theory, this seems to make the enhancement of supersymmetry very unlikely, despite the fact that all the deformed massless phases have this enhancement in the infrared. 
	
One can conceive a bizarre scenario in which our argument could fail. Note that our domain-wall calculations are only valid in the weak coupling limit, $\frac{1}{M} \ll \frac{1}{\Lambda_\text{UV}}$, where $M$ is the scale of the mass deformations. So our argument implies that supersymmetry enhancement cannot happen for {\em large} masses. However, in the Wilsonian renormalization-group picture, deformations in the UV correspond to deformations of the IR SCFT only for {\em small} masses,
and therefore one could imagine a scenario with some phase transition as we change the scale of $M$.

Alternatively, one could imagine that the enhancement occurs and the $\frac12$-BPS domain walls of the UV theory correspond to $\frac14$-BPS domain wall in the $\mc N=2$ SCFT in the IR. However, we argue that this cannot happen: In the UV, the four mass deformations corresponding to the Cartan generators of the $SU(5)$ symmetry must map to some $\mc N=2$-preserving deformations in the linear multiplets of the IR $SU(5)$. Moreover, the $SU(5)$ symmetry cannot mix with the $U(1)$ $\mc R$-symmetry, and the deformed IR theory would consequently have an $\mc R$-multiplet.  We showed in section \ref{sec:N=2 DW} this leads to $\frac12$-BPS domain walls with (2,0) or (0,2) supersymmetry.
\section{Supersymmetry enhancement from \texorpdfstring{$\mc{N}=2$}{N=2} to \texorpdfstring{$\mc{N}=4$}{N=4} \label{sec:2 to 4}}
Recently there was an interesting proposal \cite{Gang:2018huc} claiming that an $\mc{N}=2$ abelian gauge theory with a Chern-Simons term at level $-\frac32$ coupled to a chiral multiplet of unit charge has $\mc{N}=4$ supersymmetry in the IR. The goal of this section is to deform the theory and consider domain walls. In particular, we deform the $\mc{N}=2$ Lagrangian by an FI term. From the $\mc{N}=4$ point of view, we argue that this deformation corresponds to the new kind of deformation of the $\mc{N}=4$ superalgebra pointed out in \cite{Cordova:2016xhm}. We then show that domain-wall solutions of this deformed $\mc{N}=4$ algebra are necessarily $\frac14$-BPS solutions. 

Consider the Lagrangian in (\ref{N=2 Lagrangian}) with $m=0$ and  $k=-3/2$. The vacuum equations are
\beq
|\f|^2=t - \frac{3}{4 \pi}\s, \quad \s \vp=0~,
\eeq
and the model has two discrete vacua. In the first vacuum, the fields take the values $\s=0$ and  $|\f|^2=t$ and in the second $\s=\frac{4 \pi}{3}t$ and $|\f|^2=0$. As before, in a domain-wall background interpolating between these two vacua, when $Z_2>0$ there are two unbroken bulk supercharges $\bQ_+$ and $\bar \bQ_+$. 

In \cite{Gang:2018huc} it was argued that in the IR the theory has an non-Langrangian $\mc{N}=4$ description, and that the $U(1)_R$ $\mc R$-symmetry  and the $U(1)_T$ topological symmetry of the $\mc{N}=2$ model become the two diagonal $SO(2)$'s of the $SO(4)_R$ $\mc R$-symmetry of the $\mc{N}=4$ superalgebra. This implies that the multiplet of the $U(1)_T$, namely the fundamental vector-multiplet, combines with the $\mc{N}=2$ stress-tensor multiplet to give the $\mc{N}=4$ stress-tensor multiplet. Therefore, deforming the $\mc{N}=2$ theory by an FI term corresponds to deforming the emergent $\mc{N}=4$ description by a relevant deformation sitting in the stress-tensor multiplet. In \cite{Cordova:2016xhm}, it was pointed out that such deformation exists, and it is a singlet under the $\mc R$-symmetry (see appendix \ref{sec:Universal mass} for an example). Moreover, it was argued that the algebra is deformed as follows
\beq
\{ Q^{i i'}_\a, Q_\b^{j j'}\}= -\e^{ij}\e^{i'j'} P_{\a\b}+g \e_{\a\b}( \e^{ij} R^{i'j'}- \e^{i'j'} R^{ij})~, \label{deformed N=4}
\eeq
where the primed and the unprimed indices correspond to the  $SO(4)=SU(2)_R \times SU(2)_{R'}$ R-symmetries whose generators are  $R^{ij}$ and $R^{i'j'}$. The constant $g$ is the continuous parameter of the deformation. One way to argue that the $\mc{N}=4$ superalgebra is deformed in this way is as follows. Since the deformation is a singlet of the $\mc R$-symmetry, just from the index structure the only terms we can write down is the above deformation but with arbitrary coefficients in front of $R^{ij}$ and $R^{i'j'}$. Imposing the Jacobi identity, one finds that the two constants must be equal in magnitude and opposite in sign. An important feature of this algebra is that it leads to a  gapped theory as it was shown in \cite{Cordova:2016xhm}. This is in agreement with the $\mc{N}=2$ which also does not contain massless degrees of freedom.

We can further consider domain-wall backgrounds in the $\mc{N}=4$ language and deform further the algebra as
\beq
\{ Q^{i i'}_\a, Q_\b^{j j'}\}= -\e^{ij}\e^{i'j'}( P_{\a\b}+Z_{\a\b})+ g\e_{\a\b}( \e^{ij} R^{i'j'}- \e^{i'j'} R^{ij})~.
\eeq
It is more convenient to rewrite the this algebra in $SO(4)$ language
\beq
\{ Q^{I}_\a, Q_\b^{J}\}= -\d^{IJ}( P_{\a\b}+Z_{\a\b}) + g\e_{\a\b} R^{IJ}~,
\eeq
where and $I,J$ indices run from one to four, and $\e_{KLIJ}R^{IJ}$ are the generators of $SO(4)$. All operators above are Hermitian. For an arbitrary background, we can use the $SO(4)$ symmetry to bring $R^{IJ}$ in the canonical form
\beq
R^{IJ}=\begin{pmatrix} 0 & r_1 & 0 & 0 \\ -r_1 & 0  &0 & 0 \\ 0 & 0 &0 & r_2 \\ 0& 0& -r_2 &0 \end{pmatrix}~,
\eeq
and split the supercharges into two independent groups of four.
Unitarity requires that every state created by an arbitrary combination of the above supercharges needs to have non-negative norm. Since the above supercharges are Hermitian, we just need to require that the upper block and lower block of the matrix $\{ Q^{I}_\a, Q_\b^{J}\}$ are positive. As in section (\ref{sec:DM}), we consider a domain wall in the $x$-direction and go to its rest frame where $P_\m=(-m,0,0)$ and $Z_\m=(0,0,Z_2)$. In this set up, the eigenvalues of the matrix are all doubly degenerate and equal $m \pm \sqrt{ g^2 r^2 + Z_2^2}$ for $r=r_1,r_2$. Demanding that all these eigenvalues are real and positive, we arrive at the following BPS bounds
\beq\label{7bpsineq}
| g r_1 |\geq|gr_2|\geq |Z_2|, \quad m \geq \sqrt{ g^2 r_1^2 + Z_2^2}~,
\eeq
where without loss of generality we choose $r\equiv|r_1|\geq|r_2|$.
An important observation is that for non-zero $Z_2$ and $m$, 
at most one eigenvalue can be zero, namely 
$m- \sqrt{ g^2 r^2 + Z_2^2}$. This happens when $m= \sqrt2 |g r|= \sqrt2 |Z_2|$, saturating the inequalities (\ref{7bpsineq}). Hence, since each eigenvalue is doubly degenerate (except when $|r_1|=|r_2|$), we find two unbroken supercharges with the same chirality on the domain-wall background; this matches the $\mc{N}=2$ description of the theory exactly.

By analyzing the super-Poincar\'e algebra, we have found that the domain wall leaves only two unbroken supercharges. However, the situation is more subtle when one analyzes the BPS equations, which in this case are
\beq
\pa_x \s =e^2 \left( |\vp|^2 -t +\frac{3}{4 \pi} \s \right)~, ~\quad (\pa_x-iA_x)\vp= \s \vp~.
\eeq
Decomposing $\vp =e^{i \theta} |q|$ into a phase and a magnitude, the second equation implies that the phase can be gauged away by setting $A_x=\pa_x \theta$, while the magnitude satisfies
\beq
\pa_x \s =e^2 \left( |\vp|^2 -t +\frac{3}{4\pi} \s \right)~, ~\quad \pa_x |\vp|= \s |\vp| ~.
\label{BPS Yam}
\eeq
As in section \ref{sec:U(1)_0} this set of equations can be rewritten as a gradient flow problem. It is then straightforward to calculate the Morse indices at the two critical points and find $1$ and $0$. This implies that there is only one modulus, and therefore, only one massless scalar field on the domain wall, which contradicts the fact that there are two unbroken supercharges. We resolve this contradiction by arguing that there is another chiral mode on the wall coming from the gauge field.

To understand the solutions of equations (\ref{BPS Yam}) we look at the strong coupling limit $e \rightarrow \infty $ \cite{Tong:2002hi}. In the region where $\pa_x\s /e \rightarrow 0 $ the solution is
\beq
\s= \frac{4\pi}{3}( |q|^2-t)~, ~\quad |q|=\frac{\sqrt{t}}{ \sqrt{ 1-e^{\frac{-8\pi t}{3}(x-x_0)}}}~.
\eeq
Close to the wall, these expressions blow up, and therefore, are valid only for $x \gg x_0$ where the assumption $\pa_x \s /e \rightarrow 0 $ still holds. On the other side of the wall for $x < x_0$ the solution is also not valid since the square root is not real anymore. In this region we must have $|q|=0$. We can further take the limit $t \rightarrow \infty$ and the above solution implies that on each side of the domain wall the theory is at a vacuum. For $x>x_0$ the theory is Higgsed, while for $x<x_0$ the theory is in the coulomb phase with an effective Chern-Simons level after integrating out the fermions. More specifically the theory is in a $U(1)_{-1}$ gauge theory with a boundary at $x=x_0$. Because of the boundary conditions, the Chern-Simons term gives rise to a chiral edge mode \cite{Tong:2016kpv}. In fact, the condition $A_0 - u A_y=0$ leads to an edge mode moving in the $t + u \,y$ direction. In our case, the BPS equations (\ref{generalized BPS}) require $u=1$ leading to a
 $(2,0)$ algebra, which is consistent with the arguments that follow from analyzing the algebra.
\section{Summery and Discussion}
We have used domain walls as a tool for testing IR dualities and supersymmetry enhancement in the IR. Dual theories supporting domain walls are expected to exhibit the same effective theory in the IR. In certain cases, we analyzed those effective theories  and we checked some of the previously proposed dualities. In principle, the logic could also be reversed, and starting from a known duality in the bulk one can derive a new duality for the effective theories on the domain walls in one dimension less \cite{Bashmakov:2018ghn}. However, typically it is very hard to analyze the effective theory. In our examples the effective theories are free and therefore we were able to analyze them in more detail. In particular we were able to determine the amount of supersymmetry they have as well as the target space.

In the process for studying domain walls, we have also obtained some general results regarding the kinematics of $\mc N=1$ theories in three dimensions. In particular, we constructed explicitly the most general stress-tensor multiplet for $\mc N=1$ theories by reduction of the $\mc N=2$ case studied in \cite{Dumitrescu:2011iu}. 
We further wrote down explicitly the $\mc S$ superfield for several models with either $\mc N=1$ or $\mc N=2$ supersymmetry.

In section \ref{sec:5} we started with the main goal of this paper. After adding deformations, we tested the triality in figure \ref{fig:1} by matching the IR domain-wall theories. In all three cases the effective theory was a chiral $(2,0)$ sigma model with target space $\mathbb{R}\times S^1$, giving more evidence that these theories are dual. The interesting fact was that while two of these theories have $\mc N=2$ manifest supersymmetry in the UV, the third one has only $ \mc N=1$ and it was conjectured to exhibit supersymmetry enhancement in the IR. Our domain wall analysis actually showed that indeed the theory on the wall has more supersymmetry than just the one inherited from the bulk; this supports the claim that supersymmetry enhances in the IR.

Having argued that the $\mc N=1$ WZ model with $SU(3)$ global symmetry in three dimension exhibits this phenomenon of supersymmetry enhancement, one could imagine that a similar phenomenon could happen for other global symmetry groups. However, in section \ref{sec:6} we studied the WZ model with $SU(5)$ WZ symmetry and superpotential $\mathcal{W} = \frac13 \tr(\Phi^3)$, and found no supersymmetry enhancement on the domain-wall in certain phases of this theory. Hence we believe that supersymmetry enhancement is very unlikely in this case.

Finally in section \ref{sec:2 to 4}, we tested the $\mc{N}=4$ supersymmetry enhancement of the $\mc{N}=2$ $U(1)_{-\frac32}$ theory coupled to a chiral multiplet of unit charge, which was conjectured in \cite{Gang:2018huc}. This theory displays two interesting phenomena. First, the FI term in the $\mc N=2$ description corresponds to a deformation of the $\mc N=4$ language by the $\mc R$-symmetry generator, recently discussed in \cite{Cordova:2016emh}. Second, domain walls in this theory interpolate between vacua of different Chern-Simons levels. As a result there is one more zero mode coming from broken gauge symmetry on the wall--this is similar to the quantum Hall-effect.

In conclusion, we have used domain walls as a non-perturbative tool for analyzing certain aspects of the infrared dynamics of strongly coupled theories. Although the phase diagrams already contain a lot of information about the theory, in phases with degenerate vacua, there exists information which can be extracted by studying domain walls. In particular, domain walls can be used to test various proposed dualities, and in supersymmetric cases, they can be used to study the possibility of supersymmetry enhancement in the IR.

\section*{Acknowledgments}
We are happy to thank Zohar Komargodski for his many crucial insights and encouragement. KR would like to thank J.P. Ang for many fruitful discussions. The work of KR and MR is supported in part by the NSF Grant PHY-1620628. The work of SS is supported by Simons Foundation Grant 1150287-1-79154.
\section*{Appendices}
\appendix
\section{Conventions and Identities \label{sec:conventions}}
\subsection{Indices and \texorpdfstring{$\g$}{gamma}-matrices}\label{indices}
We use a $(-,+,+)$ signature. The conventions for different kind of indices are

\begin{itemize}
\item $\a,\b,\g \dots$ are flat spinor 
\item $a, b,c \dots$ are curved spinor 
\item $\m, \n, \r \dots$ are flat vector
\item $M,N,P \dots$ are flat superspace
\item $A,B,C \dots$ are curved superspace
\item $i,j,k \dots$ are fundamental flavor
\item $I,J,K \dots$ are adjoint flavor
\item $\mm, \nn, \pp \dots$ are fundamental color
\item $\MM, \NN, \PP \dots$ are adjoint color
\end{itemize}
We raise and lower spinor indices with the epsilon symbol where $\e_{+-}=-1=-\e^{+-}$
\beq
\psi^\a=\e^{\a\b} \psi_\b, \quad \psi_\a=\e_{\a\b} \psi^\b~.
\eeq
For a product of two spinor variables we have 
\beq
\psi \chi= \psi^\a \chi_\a=-\psi_\a \chi^a~.
\eeq 
The gamma matrices are taken to be $(\g^\m)_\a{}^\b:= \e^{\b\d}\g^\m_{\a\d}$
with
\beq
\g^\m_{\a\b}=(-\mathbbm{1}, \s_1,\s_3 )~, 
\eeq
where $\s_i$ are the usual Pauli matrices. They satisfy
\bea
& \g^\m_{\a\b}=\g^\m_{\b\a} ~,\\
& (\g^\m)_\a{}^\b (\g^\n)_\b{}^\g = \eta^{\m\n} \d_\a{}^\g + \e^{\m\n\r} (\g_\r)_\a{}^\g ~, \\
& (\g^\m)_{\a\b} (\g_\m)_{\g\d}=\e_{\a\g}\e_{\d\b}+\e_{\a\d}\e_{\g\b}~.
\end{align}
Here $\e_{012}=-1$.
A vector with two spinor indices is defined as
\beq
l_{\a\b}=-2 \g^\m_{\a\b} l_\m, \quad l_\m=\frac{1}{4}\g_\m^{\a\b} l_{\a\b}~. 
\eeq 
\subsection{Superspace}\label{susy_A}
Derivatives with respect to $\theta$ variables are defined by
\bea
\pa_\a \theta^\b=\d_\a{}^\b ~, \\
\pa^\a \theta^\b=\e^{\a\b} ~,\\
\pa_\a \theta_\b=-\e_{\a\b}~. 
\end{align} 
Our conventions regarding the Hermiticity properties of various quantities are 
\beq
(\theta^a)^\dagger= \theta^a, \quad (x^{\a\b})^\dagger=x^{\a\b}~.
\eeq
From 
\beq
\{ \pa_\a, \theta^\b \}= \d_\a{}^\b, \quad [\pa_{\a\b}, x^{\g\d} ]= -4 \d_{(a}{}^\g \d_{\b)}{}^\d~,
\eeq
it follows that, as operators, 
\beq
(\pa_\a)^\dagger= \pa_\a, \quad (\pa_{\a\b})^\dagger=-\pa_{\a\b}~.
\eeq
The supercharges acting on superfields are given by
\beq
Q_\a= \pa_\a- \frac{i}{2} \theta_\b \pa_\a{}^\b~.
\eeq
and they satisfy 
\beq
\{Q_\a,Q_\b \}=i\pa_{\a\b}=-2 i \g^\m_{\a\b} \pa_\m=2 \g^\m_{\a\b} P_\m ~,
\eeq
and have the following hermiticity properties
\beq
(Q_\a)^\dagger= Q_\a~.
\eeq
We furthermore define how conjugation acts on a product
\beq
( A B )^\dagger= B^\dagger A^\dagger~.
\eeq
In our conventions the gamma matrices are real, and therefore, all the fermionic fields are real as well. Integration with respect to the anticommuting coordinates is defined by
\beq
\int d \theta_\a= \frac{1}{2}\pa_\a~.
\eeq
To make the Langragian (\ref{WZ Lagrangian}) of the WZ model real, note that we have to include an unusual factor of $i$ in from of the superpotential. Covariant derivatives for $\mc{N}=1$ superspace are taken to be
\bea
&\{D_\a, D_\b\}=-i \pa_{\a\b}, \\
& D_\a=\pa_\a-i (\g^\m)_\a{}^\b \theta_\b \pa_\m=\pa_\a+\frac{i}{2} \theta_\b \pa_\a{}^\b ~,
\end{align}
and satisfy the following useful identities
\bea
& D_\a D_\b= -\frac{i}{2} \pa_{\a\b} +\frac12\e_{\a\b} D^2~ \\
& D^2 D_\a=-D_\a D^2 ~,\\
& D^\a D_\b D_\a =0 ~,\\
& \pa^{\a\b}D_{\a}D_{\g}D_{\b}=0 ~. \label{pDDD}
\end{align}
We define a real scalar superfiled $\Phi(x,\theta)$; its components are
\beq
\Phi|=\vp~,~~D_\a\Phi|=i\psi_\a~,~~D^2 \Phi|=iF~.
\eeq

Superspace for $\mc{N}=2$ supersymmetry can be built out of two copies of $\mc{N}=1$ superspace as 
\begin{equation}
\theta_\a=\frac{1}{\sqrt{2}}  \left( \theta_\a^{(1)} +i \theta_\a^{(2)} \right)~, \quad
\bar{\theta}_\a=\frac{1}{\sqrt{2}}  \left( \theta_\a^{(1)} -i \theta_\a^{(2)} \right)~.
\end{equation}
This definition leads to the following expression for the covariant derivatives
\begin{equation}
{\bD}_\a  =\frac{1}{\sqrt{2}}  \left( D_\a^{(1)} -i D_\a^{(2)} \right)~, \quad
\bar{{\bD}}_\a  = -\frac{1}{\sqrt{2}}  \left( D_\a^{(1)} +i D_\a^{(2)} \right)~,
\end{equation}
which satisfy
\bea
& \{{\bD}_\a,\bar{{\bD}}_\b \}=i \pa_{\a\b}~, \nonumber \\
& \{{\bD}_\a,{\bD}_\b \}=\{\bar{{\bD}}_\a,\bar{{\bD}}_\b \}=0~.
\end{align}
In terms of derivatives there are given by
\begin{equation}
{\bD}_\a  = \pa_\a+\frac{i}{2} \bar{\theta}_\b \pa_\a\!^\b~, \quad \bar{{\bD}}_\a  =-\bar{\pa}_\a-\frac{i}{2} \theta_\b \pa_\a\!^\b ~.
\end{equation}
Some very useful identities \cite{Zupnik:1999tf} are
\bea
{\bD}_\a {\bD}_\b & = \frac{1}{2} \e_{\a\b} {\bD}^2~, \quad {\bD}_\a {\bD}_\b {\bD}_\g  =0 ~,\\
[\bar{{\bD}}_\a, {\bD}^2 ] & = 2 i \pa_\a\!^\b {\bD}_\b~, \quad  
{\bD}^2 \bar{{\bD}}^2 {\bD}^2= 4 i {\bD}^2  \Box ~, \\
{\bD}_\a \bar{{\bD}}_\b & = \frac{i}{2} \pa_{\a\b} 
+\frac{1}{4}[{\bD}_{(\a},\bar{{\bD}}_{\b)}] +\frac{1}{2} \e_{\a\b} {\bD} \cdot \bar{{\bD}}~, \\
[{\bD}_{(\a},\bar{{\bD}}_{\b)}]\,  {\bD}\cdot \bar{{\bD}} & = \frac{i}{2} \left( \pa_\a \!^\g  [{\bD}_{(\g},\bar{{\bD}}_{\b)}] + \pa_\b \!^\g  [{\bD}_{(\g}~,\bar{{\bD}}_{\a)}] \right)~.
\end{align}
\subsection{Super Yang-Mills theory}\label{sym_A}
Now we define the gauge-covariant superderivatives for $\mc{N}=2$ gauge theories following \cite{Hitchin:1986ea}. They satisfy the following algebra
\bea
\{ \na_\a, \na_\b \} &= \{ \bar{\na}_\a, \bar{\na}_\b \} = 0~, \label{constraints2}\\
\{ \na_\a, \bar{\na}_\b \} &= i\na_{\a \b} + i \e_{\a \b} \Sigma~.
\end{align}
A particular useful solution to constraints \eqref{constraints2} is the chiral representation,
\bea
\na_{\a} = e^{-V} {\bD}_\a e^V, \quad \bar{\na}_\a = \bar{{\bD}}_\a~.
\end{align}
$\Sigma$ is the scalar field strength and in this representation,
\beq
\Sigma = \frac{i}{2} \e^{\a \b} \{ \na_\a, \bar{\na}_\b \}
= \frac{i}{2} \bar{{\bD}}^\a (e^{-V} {\bD}_\a e^V)~.
\eeq
Now using the Bianchi identity, we find
\bea
[\bar{\na}_\a, \{ \bar{\na}_\b, \na_\g \} ] &= +
2i  \e_{\a \b}\bar{\na}_\g \Sigma ~,\\
[\na_\a, \{ \na_\b, \bar{\na}_\g \} ] &=
-2i  \e_{\a \b}{\na}_\g \Sigma ~,\\
[\bar{\na}_{\a}, \na_{\b \g}] &=+ \e_{\a (\b} \bar{\na}_{\g)}\Sigma~,\\
[\na_{\a}, \na_{\b \g}] &= -\e_{\a (\b} \na_{\g)}\Sigma~.
\end{align}
Thus for the vector covariant derivatives $\na_\mu = -\frac{i}{4} \g_\mu^{\a \b} \{ \na_\a, \bar{\na}_\b \}$, we have
\bea
[\na_{\a},\na_{\mu}] &= \frac{1}{2} {(\g_\mu)}_{\a \b} \na^\b \Sigma~,\\
[\bar{\na}_{\a},\na_{\mu}] &= -\frac{1}{2} {(\g_\mu)}_{\a \b} \bar{\na}^\b \Sigma~.
\end{align}
By using the Bianchi identity $\{\na_{(\a}, [\na_{\b)},\na_{\mu}] \} = 0$, we find that $\Sigma$ is covariantly linear $\na^2 \Sigma = \bar{\na}^2 \Sigma = 0$. Finally, we get the non-trivial part of the full algebra as
\bea
\{ \na_\a, \na_\b \} &= \{ \bar{\na}_\a, \bar{\na}_\b \} = 0~, \label{constraints}\\
\{ \na_\a, \bar{\na}_\b \} &= i\na_{\a \b} + i \e_{\a \b} \Sigma~,\\
[\na_{\a},\na_{\mu}] &= \frac{1}{2} {(\g_\mu)}_{\a \b} \na^\b \Sigma~,\\
[\na_\mu, \na_\nu] &= -i {\bF}_{\mu \nu} = -i \e_{\m \nu \rho} \tilde{{\bF}}^\rho~,
\end{align}
where
\beq
\tilde{{\bF}}_{\a \b}
= \frac{1}{4} \left[ \na_{(\a} , \bar{\na}_{\b)} \right] \Sigma, \quad \text{and} \quad \na_\mu {\Big |} = \pa_\mu - i A_\m~.
\eeq
For the chiral representation, under the complexified gauge transformations we have
\begin{gather}
\Phi \rightarrow e^{i\Lambda} \Phi, \quad \bar{\Phi} \rightarrow \bar{\Phi}e^{-i\bar{\Lambda}}~,\\
\na_{M} \rightarrow e^{i\Lambda} \na_{M} e^{-i\Lambda}, \quad \bar{\na}_{\a} \rightarrow \bar{\na}_{\a}~,\\
e^V \rightarrow e^{i\bar{\Lambda}} e^V e^{-i\Lambda} \quad\Sigma \rightarrow e^{i\Lambda} \Sigma e^{-i\Lambda}~,
\end{gather}
where $\Lambda=\Lambda^A T_A$ is a chiral multiplet and $\na_M=(\na_\a,\na_\mu)$~.

We define the covariant components of the scalar field strength as
\begin{gather}
\Sigma {\Big |} = 2\s, \quad \na_\a \Sigma {\Big |} = 2\bar{\l}_\a, \quad \bar{\na}_\a \Sigma {\Big |} = 2\l_\a,\\
\na \cdot \bar{\na} \Sigma {\Big |} = \bar{\na} \cdot \na \Sigma {\Big |} = -4i D~, \quad
\frac{1}{2} \left[ \na_{(\a} , \bar{\na}_{\b)} \right] \Sigma {\Big |} = 2\tilde{{\bF}}_{\a \b} {\Big |} = 2\tilde{F}_{\a \b}~,
\end{gather}
and the covariant components of the covariantly chiral (anti-chiral) superfields $\Phi$ and $\bar\Phi e^V$ as:
\begin{gather}
\Phi|=\vp~,~~~~\na_\a\Phi|=i\psi_\a~,~~~~\na^2 \Phi|=iF~,\\
\bar\Phi e^V{\big|}=\bar\vp~,~~\bar\na_\a\bar\Phi e^V{\big|}=-i\bar\psi_\a~,~~\bar\na^2 \bar\Phi e^V{\big|}=i\bar F~.
\end{gather}
\section{Three-dimensional \texorpdfstring{$\mc{N}=1$}{N=1} Supergravity \label{sec:sugra}}
Following \cite{Gates:1983nr}, an $\mc{N}=1$ theory in three dimension couples to supergravity in a trivial way
\beq
L=\int d^2 \theta \; f(\phi, D_A \phi) \rightarrow  \int d^2 \theta \; E^{-1}  \; f(\phi, \na_A \phi)~,
\eeq
where $E_A{}^M$ is the super-vielbein and $\na_A$ is the super-covariant derivative
\beq
\na_M= E_M{}^A D_A +  \phi_{M \b}{}^\g M_\g{}^\b~,
\eeq
with $\phi_{M\b}{}^\g$ a connection and $M_\b{}^\g$ the local Lorentz generators.
These transform under super-diffeomorphisms $K=K^AD_A$ and local Lorentz transformations 
$L=L_\a{}^\b M_\b{}^\a$ as
\beq\label{a.b.trans}
\d \na_M=[ \na_M,K+L]~.
\eeq
The super-torsion and the super-Riemann tensor are defined by
\beq
 [ \na_M, \na_N \}= T_{M N}{}^{R} \na_R + (R_{MN })_\a{}^\b M_\b{}^\a~.
\eeq
Both of them can be expressed in terms of the connection and the super-vielbein. 

Imposing the conventional constraint 
\beq
\na_{\a\b}= - \frac{i}{2} \{\na_\a, \na_\b\}~,
\eeq
we fix two components of the super-vielbein, namely
\bea
 E_{\a\b}{}^a & = E_{(\a}{}^A D_A E_{\b)}{}^a - i \phi_{\a\b}{}^\g E_\g{}^a ~, \\
 E_{\a\b}{}^{a b} & =  E_{(\a}{}^A D_A E_{\b)}{}^{ab} + \frac{1}{2} E_\a{}^{(a} E_\b{}^{b)} -i  \phi_{\a \b}{}^\g E_\g{}^{a b}~.
\end{align}
For simplicity we focus only at the linearized theory with
\bea
E_\a{}^a & = \d_\a{}^a +e_\a{}^a~, \\
E_{\a\b}{}^{ab} & = \frac{1}{2} \d_\a{}^{(a} \d_\b{}^{b )} +e_{\a\b}{}^{a b} ~,\\
E_\a{}^{ a b} &= e_\a{}^{ab} ~,\\
E_{\a\b}{}^a &= e_{\a\b}{}^a~.
\end{align}
A generic theory will couple to the two independent components of the super-vielbein by
\beq
\int \left( e_\a{}^\b J_\b{}^\a + e_\g{}^{\a\b}  J_{\a\b}{}^\g  \right)~.
\eeq
where now all the indices are flat and the $J$'s are some currents. Invariance under the linearized versions of (\ref{a.b.trans}) 
\bea
\d e_\a{}^\b & = D_\a K^\b - L_\a{}^\b~, \\
\d e_\a{}^{\b\g} & = D_\a K^{\b\g} -i  \d_\a^{(\b} K^{\g)}~.
\end{align}
implies that the $J$'s obey the following equations
\bea
D_\a J_{\b\g}{}^\a & =0~, \\
D^\b J_{\a\b} & =- J_{\a \b}{}^\b~, \\
J_{\a\b} & =   \e_{\a\b}  J~,
\end{align}
where the last equation follows because the Lorentz parameter $L_{\a\b}$ is symmetric. Equivalently we have
\begin{align}
D^\b J_{\a\b,\g} & = -D_\g D_\a J, \nonumber \\
J_{\a\b}\: ^\b & = - D_\a J~,
\end{align}
which match (\ref{D2 on phi}) under the identification $\mcs_{\a\b,\g}=J_{\a\b,\g}$ and $X=-J$.

\section{Deformations of a free Hypermultiplet \label{sec:Universal mass}}
In three dimensions, the $\mc{R}$-symmetry group of $\mc{N}=4$ theories is $SO(4)=SU(2)_R\times SU(2)_{R'}$. A free hypermultipet in three dimensions consist of four scalar and four fermionic fields and has an additional $SU(2)_F$ flavor symmetry. The supersymmetric transformations are simply
\beq
\d \vp_i^a= \e^\a_{i i'} \psi^{i'a}_\a, \quad \d\psi^{i' a}_\a= \e^{\b,i i'}\pa_{\a\b}\vp_i^a~.
\eeq 
As in the main text $i$ and $i'$ are indices or $SU(2)_{R}$ and $SU(2)_{R'}$ respectively, while $a$ is an  $SU(2)_F$ index. As was pointed out in \cite{Cordova:2016emh}, apart from F-terms, the model has two relevant deformations. The first belongs to the supermultiplet of the $SU(2)_F$ which in the notation of \cite{Cordova:2016emh} is the $B_1 [0]_1^{(2,0)} $ multiplet with lowest component $\bar{\vp}^i_a \vp^a_i$. Acting twice with the above supersymmetric transformations, we get the conserved flavor current plus the scalar $\bar{\psi}^{(i'}{}_{(a} \psi^{j')}{}_{b)}$. Furthermore, the transformation of this scalar is a total derivative after we use the equations of motion. Deforming the theory with this term corresponds to adding a real mass to the fermions. However, since we have already used the equations of motion to show that the supersymmetric transformation of this quantity is a total derivative, in principle, we have to modify the transformations as well as to add more terms to the Lagrangian that are higher order terms in the deformation constant. Hence
\beq
\d \mc{L}= m_{i' j'}^{ab} \bar{\psi}^{(i'}{}_{(a} \psi^{j')}{}_{b)}  + \mc{O} (m^2)~.
\eeq
Moreover, in this simple case we already know what the higher order terms in $m$ are, and how the transformations change. The additional term is of course the mass term for scalars, namely  $m_{i' j'}^{ab} m^{i' j'}_{ab}  \vp^c_i \vp_c^i$, while the new transformations are 
\beq
\d \vp_i^a= \e^\a_{i i'} \psi^{i'a}_\a, \quad \d\psi^{i' a}_\a= \e^{\b,i i'}\pa_{\a\b}\vp_i^a + \e^\b_{ij'} (m^{ab})^{i'j'} \vp_b^i~.
\eeq 
The second relevant deformations belong in the $\mcs$-multiplet of the $\mc{N}=4$ model, whose lowest component is $\vp^a_i \vp_a^i$. In the notation of \cite{Cordova:2016emh} this is the $A_2[0]_1^{(0,0)}$ multiplet. In this case the deformation is an singlet of the $\mc{R}$-symmetry, and repeating the above logic we arrive at the following deformation
\beq
\d \mc{L}= m \bar{\psi}_a^{i'}  \psi_{i'}^a  + m^2 \vp^c_i \vp_c^i~,
\eeq
with transformations
\beq
\d \vp_i^a= \e^\a_{i i'} \psi^{i'a}_\a, \quad \d\psi^{i' a}_\a= \e^{\b,i i'}\pa_{\a\b}\vp_i^a +m  \e_\b^{ii'}\vp^a_i~.
\eeq 

\bibliographystyle{JHEP}
\bibliography{refs}

\end{document}